# Form Follows Function: A Method of Discovering How Neurons Are Connected To Process Information

Lane Yoder
Department of Science and Mathematics, retired
University of Hawaii, Kapiolani
Honolulu, Hawaii
lyoder@hawaii.edu
NeuralNanoNetworks.com

**Abstract**

A method of discovering how neurons are connected to process information is presented here: Design a simple logic circuit that can perform a single, biologically advantageous function. Engineering concepts can be helpful in choosing the function and in designing the logic circuit. Several implementations of the method are reviewed to demonstrate how a biologically advantageous function can be chosen, how one simple network can generate major phenomena that are widely considered unrelated, and how one network design can lead to others that explain entirely different aspects of the brain. These results show that the method can benefit neuromorphic engineering as well as neuroscience, and that some brain functions can be carried out remarkably simply, at least in principle if not in the details.

The network models produce dozens of known phenomena and predict unknown phenomena. Apparently none of the phenomena had been explained previously. Other than the single function for which each model was designed, all of the phenomena produced by the model are by-products that have no significance in the function that the model was designed to produce. Known phenomena that are by-products of the model are crucial for showing how neurons are actually connected.

Retinal cones encode spectral information in a photostimulus. A 13-neuron decoder with input from three classes of cones was designed only for consistent discrimination of spectral distributions under different lighting intensities. The analog (fuzzy logic) decoder generates a dozen phenomena central to local color vision, including color vision's "shortcomings." An extension of the fuzzy decoder design to allow any number of inputs produces olfactory identification of odorants independently of their concentrations and the major phenomena of olfaction. With one additional neuron, the extended decoder makes up a family of general information processors that can produce all of the brain's combinational processing of information. These networks exhibit major features of the cerebral cortex's physiology and anatomy.

Flip-flops are the basic building blocks for sequential processing of information. A flip-flop with digital (Boolean logic) outputs, designed to store one bit of information, can be formed with as few as two neurons. One of the neurons in a flip-flop exhibits the seven characteristics associated with short-term memory formation, retention, retrieval, termination, and errors. Cascaded oscillators, consisting of a three-neuron ring oscillator connected in sequence with flip-flop toggles, were designed with a variety of speeds to synchronize state changes in neural structures to avoid timing errors. The synchronized state changes accurately produce the multimodal distribution of brain wave frequencies in the usual bands. The model correctly predicts a previously unknown relationship: The entire distribution of brain wave frequencies is an explicit function of just two parameters - the mean and variance of neuron delay times. The cascade design's necessarily narrow tolerance for error suggests a possible relationship to the abnormal electrical activity characteristic of epileptic seizures. Together, flip-flop memory mechanisms and synchronization by flip-flop oscillators suggest a resolution to the longstanding controversy of whether short-term memory depends on neurons firing persistently or in brief, coordinated bursts. Flip-flop oscillators show how two primitive ganglia that have been studied extensively can produce lobster peristaltic action and lamprey locomotion.

For each network design, all neurons, connections, and types of connections are shown explicitly. The neurons' operation depends only on explicitly stated, minimal properties of excitement and inhibition. This operation is dynamic in the sense that that only the strengths of the neurons' signals are subject to change, depending on the current strengths of the neurons' inputs, making the networks' operation consistent with the speed of most brain functions. The design method includes an element of rigor: Conclusions that the networks can generate neural correlates of certain phenomena can be proven from the models' explicit architectures and minimal neuron capabilities. This holds for both known phenomena and testable predictions of unknown phenomena. Together, the models include examples of every kind of information processing: analog and digital, combinational and sequential, parallel and serial.

Since the designs are logic circuits, they can be implemented with electronic components. Some of the designs discussed here are apparently new to engineering, filling gaps and providing improvements in well-known families of logic circuits. A simple transformation can convert certain electronic logic circuit designs to network designs that can be implemented with neurons, and vice versa. This transform was used to derive a neuronal flip-flop from a standard electronic design, to derive a novel electronic oscillator from a proposed neuronal model for the lobster STG, and to show that the known neuron connectivity that controls the lamprey's swimming motion incorporates the most popular electronic flip-flop design.

## 1. Introduction

How the brain functions is widely considered one of the most important unanswered questions in biology. Although neurons are varied and complex, at the most basic level their signaling method is rather simple and the same in all animals: A neuron excites or inhibits other neurons. How neurons process information is determined mainly by the organization of synaptic connections. Besides the number of neurons, connectivity is the main feature that distinguishes the brain between species and even between individuals. This is the thesis of Seung's classic *Connectome* [1].

The relationship between a neuronal network's organization of synaptic connections and how the network produces its results is poorly understood. Even in primitive brains and simple ganglia whose neural "wiring" has been fully mapped, how the networks produce their results is not known. This is one of the main points in the introduction to Churchland and Sejnowski's *The Computational Brain* [2, pp. 4, 5].

The form follows function (FFF) method - designing a logic circuit that can perform a single, biologically useful function - suggests a possible way forward. The process of using the method is discussed here, demonstrated by several implementations of the method [3-10]. These articles will be referred to as the FFF papers. In the interest of brevity, only demonstrations of the diverse ways the method has been applied and the main results are presented here. For more thorough explanations and supporting evidence for the results, specific references are given to the FFF papers. Several novel observations, discoveries, and predictions that were not mentioned in the abstract are discussed in the Conclusions section.

The networks in the FFF papers include the two fundamental kinds of information processing. Papers [3-6] present decoders. These networks can be configured as general information processors that can produce all of the brain's combinational processing of information, i.e., logic operations whose outputs depend only on the current state of the inputs. Papers [7-10] explore applications of flip-flops, the basic building blocks for sequential processing of information. This means logic operations whose outputs are functions of both the current inputs and the past sequence of inputs.

All of the conclusions that the networks can generate neural correlates of certain phenomena can be proven from the models' explicit architectures and explicitly-stated, minimal neuron capabilities. Some of the network results are demonstrated by simulation, but the results are not dependent on the simulations.

The most surprising aspect of the FFF method may be the number and variety of major phenomena that can arise from a single network design. A prominent neuroscientist who saw the first draft of this manuscript emailed about the decoder, "…nobody else thinks that all those diverse phenomena boil down to the same kind of interaction. Each is thought to be its own, unique, challenge."

One of the main benefits of a logic circuit design by "form follows function" is that if the chosen function and theoretical design both match evolution, then the design will also match evolution in whatever additional phenomena are generated as by-products of the design's intended function. For the networks found in the FFF papers, each of the several phenomena produced by a single network design is a potentially falsifying prediction. Both the number of known phenomena produced by each FFF network and their independence, i.e., none of them implies any other, make the phenomena likely to be produced by the proposed theoretical network in the actual nervous system. The variety of brain functions generated by minor modifications of a single design may be examples of evolution finding new uses for what it already has.

The main novel contribution of this paper is the FFF method of discovering how neurons are connected to process information. The method was not mentioned in any of the FFF papers, and apparently it has not been proposed elsewhere in the literature. The FFF papers are reviewed here to demonstrate how the method can be applied and what kinds of results can be expected. Although much of the review was necessarily taken from the FFF papers, several parts have not been published previously. Significant aspects of the review are the discussions of how the models in the different FFF papers are related to each other, how to choose an advantageous function to model, and what characteristics the models' have in common (discussed mostly in section 3 and demonstrated by the examples throughout the paper). None of this essential information could be readily inferred from the FFF papers. Other novel elements of this article that did not appear in the FFF papers: Section 2 on other approaches to neuron connectivity; the explanation in the beginning of section 4.4 of why one of the models exhibits major features of the cerebral cortex; the new models for neural correlates of the perceptions of violet and purple in Figs 10A and 10B; and Figs 2, 3, 6, 7, and 10.

The review of the literature in this article is unusually short. The subject of the article is a method of discovering how neurons are connected to process information in the real time of most brain functions. There are no other methods that even claim to do this, let alone accomplish it. All of the methods that consider connectivity either do not show explicit connections or do not show how the networks can produce known results. This point is discussed further in sections 2 and 3.2.

## 2. Other approaches to neuron connectivity

Several different methods used in attempts to discover how neurons are connected have had little success in finding how a specific organization of synaptic connections produces a network's results.

### 2.1. Direct inspection

Some of the earliest approaches to neuron connectivity involved direct inspection by microscope, à la Camillo Golgi and Ramón y Cajal. The most common modern brain imaging techniques are functional magnetic resonance imaging (fMRI), computerized tomography (CT), positron emission tomography (PET) and Magnetoencephalography (MEG). While these methods have been successful in finding where various kinds of information are processed and how these regions are connected, none of the modern imaging techniques has resolution high enough to detect connections between neurons. So the technology cannot be used to explain how individual neurons are connected to process information.

Electron microscopy has been more successful in small-scale imaging. Complete "wiring" maps have been found for a few small nervous systems and ganglia, such as the fruit fly brain (thousands of neurons), flatworm brain (300), lobster gastric mill (30), and lamprey locomotion (eight-neuron segmental ganglia). Much has been learned from such maps, but except for the lamprey's ganglion in one of the FFF papers [10], these efforts have famously failed to show how the specific connections produce their results. "Boiled down, the lesson is that microlevel data are *necessary* to understand the system, but not *sufficient*." [2, p. 5].

Most recently (Oct 3, 2024), a multinational consortium of 287 researchers at 76 institutions published a complete map of an adult fruit fly brain, including all 139,255 neurons and all connections (~50 million) [11]. In an interview of several co-authors [12], one of them acknowledged that while this kind of research shows "interesting network structure we would very much like to understand in greater detail," it does not address the "central question" of how networks of neurons are organized to "process information, drive behavior, and store and recall memory." The author added, "Figuring this out is a hard problem…."

### 2.2. Design a neuronal network that can produce known phenomena

The few attempts at designing an explicit network to produce known phenomena have met with little success. The problem may be that a network that performs just one function can produce several phenomena as by-products that seem to be unrelated. This can lead to proposed network designs that essentially include a separate network for each phenomenon.

For example, a well-known two-stage color vision model [13, 14] proposed that a neuron with input from photoreceptors that are sensitive to middle and long wavelengths can mediate the perception of yellow. However, unless the neuron can discriminate the relative strengths of the inputs, it cannot distinguish between yellow and red or green. The model also cannot account for the reddish appearance of the shortest wavelengths. Later a four-stage model [15] attempted to deal with these problems, adding more complexity.

A few more recent attempts using this method were discussed in some detail in [7, section 2.3], but each one has significant problems.

Some researchers have suggested that the models in the FFF papers were simply designed to produce known brain phenomena. That was not the case. For most of the phenomena, I was not aware that the phenomenon was known to exist or not aware that the model produced it (or both) until after I had designed the model. Given an arbitrary set of several complex phenomena, a method for designing a single mechanism that can produce all of them is not obvious. If some of the phenomena have separate causes, such a mechanism is unlikely even to be possible.

### 2.3. Neuromorphic engineering and computing

AI and machine learning have been extraordinarily successful, but the approach has been more to apply presumed neuronal networks to AI rather than to discover how nervous systems function. Moreover, the neuromorphic approach suffers from a lack of “an accurate neuroscience model of how the brain works” [16].

### 2.4. Neural engineering

The results in the FFF papers suggest that both neuroscience and engineering can benefit from the opposite of neuromorphic engineering, i.e., neuroscience derived from engineering concepts. The field of neural engineering is relatively new. The first journals devoted to the subject began in 2004. Neural engineers have applied engineering concepts to study a wide variety of aspects of the brain, but curiously these have apparently not included how neurons are connected to process information. This is especially surprising because so much is known about how electronic components can be connected to process information.

### 2.5. Large projects

In recent years there has been lavish support for national and multinational research projects on the brain (e.g., the EU Human Brain Project, the US BRAIN Initiative, the Japan Brain/MIND project). These projects have included such massive experiments as whole brain simulation on supercomputers and in vitro growth of large colonies of neurons to see what happens. The experiments have produced results, but apparently not in finding how a network's specific organization of synaptic connections produces the network's results. These efforts have received criticism even within the scientific community for their excessive largesse at the expense of other research, lack of direction and control, and again, lack of progress.

## 3. Method: Form follows function

The FFF method presented here for discovering neuron connectivity is to design a simple logic circuit that can perform a single, biologically advantageous function. The process of choosing an advantageous function and the characteristics of the results are discussed in this section.

### 3.1. Choosing a single, biologically advantageous function

In applying the FFF method, the main challenge is choosing a useful function out of the apparent myriad of brain functions. The examples in the FFF papers show that in some cases the right choice is much more obvious than in others, and that engineering concepts can be helpful in making both the choice of function and the logic circuit design. One elementary course in logic circuit design is more than enough to produce the results in the FFF papers.

### 3.1.1. Color vision from two problems

A simple function that might contribute to color vision is not obvious. There are two major problems in discriminating spectral distributions in different photostimuli: All photoreceptors are somewhat sensitive to all parts of the visible spectrum, and the intensity of natural light varies throughout the day and when the weather or shelter changes.

The color vision model proposed in [3] and refined in [4] was discovered by considering a simplified case for these problems. Suppose a primitive sea creature has two photoreceptors, one that is mostly sensitive to longer wavelengths and another that is mostly sensitive to shorter wavelengths. The two receptors by themselves are not enough to discriminate between long and short wavelengths because both cells can respond to both wavelengths. There would be a need for a way of comparing the two responses to distinguish a photostimulus of solar light reflected from a yellow object (i.e., long wavelengths) from the blue light scattered by the surrounding water (short wavelengths). A neuron can provide the comparison because it sends a signal when it has greater excitatory input than inhibitory input. A neuron with excitatory and inhibitory inputs from the long and short wavelength receptors, respectively, could produce the distinction independently of the intensity of solar light.

The rest of the color vision decoder model was derived by extension of this little thought experiment, described further below.

### 3.1.2. Olfaction by extension

Olfaction faces problems that are analogous to the color vision problems: overlapping olfactory receptor neuron sensitivities to different chemicals and varying odorant concentrations.

An extension of the color decoder design to allow any number of inputs can identify odorants independently of concentrations [4]. The extended decoder also produces olfaction's major phenomena, which are analogous to the color vision phenomena.

### 3.1.3. Cerebral cortex physiology and anatomy by observation

Perhaps most surprisingly, the cerebral cortex results show how powerful the FFF method can be. No biologically advantageous function was considered in this case. It was simply observed that the extended decoders exhibit several of the major features of the cerebral cortex [5]. There is a plausible explanation for this startling result that was not included in [5], and it is given in more detail below: With the minor modification of adding one cell, the extended decoder makes up a family of general information processors that can produce all of the brain's combinational processing of information, i.e., logic operations whose outputs depend only on the current state of the inputs.

### 3.1.4. Short-term memory from logic circuit design

Memory is clearly a useful function, and the fast formation of short-term memory is especially advantageous. The engineering field of logic circuit design has covered different types of memory rather thoroughly. The flip-flop is the standard dynamic memory mechanism for one bit of memory in electronic computational systems. A two-neuron flip-flop (or latch), discussed below, was derived directly from an electronic design by the transform mentioned in the abstract, also discussed below. Neuronal flip-flops can generate known phenomena of short-term memory formation, retention, retrieval, termination, and errors [7]. Flip-flops are also the basic building blocks for sequential processing of information, i.e., logic operations whose outputs are functions of both the current inputs and the past sequence of inputs.

### 3.1.5. Brain waves from synchronization to avoid timing errors

A useful function that can produce the phenomena of brain waves may be the least obvious because the function of brain waves is itself uncertain. A well-known engineering concept provided a partial answer here. State changes in nearly all electronic computational systems are synchronized by an oscillator, or clock, to avoid timing errors. Timing problems are greater in sequential logic than in combinational logic, and greater in parallel processing than in serial processing. Much of the information processing in the brain involves sequential logic and nearly all of it is parallel. This means the selective pressure for synchronization of state changes of neural structures in the brain would have been high, and the neural implementation is quite simple.

The next step in usefulness is that a large system like the brain would need a wide variety of frequencies for different types of information processing, including some information that needs to be processed as fast as possible. This requirement leads directly to cascaded toggles functioning as oscillators [8], another well-known engineering concept. A neuron can function as an inverter, and a three-inverter ring oscillator is the fastest oscillator that can be constructed with logic gates. The ring oscillator connected in sequence with flip-flop toggles makes the toggles oscillate. The toggles provide a wide variety of speeds because the oscillation period grows exponentially (doubling) with each successive oscillator. State changes synchronized by such oscillators accurately produce the somewhat complex multimodal distribution of brain wave frequencies in the usual bands [8].

### 3.1.6. Lobster STG and lamprey locomotion from oscillator concepts

The functions of the lobster's stomatogastric ganglion (STG) and the lamprey's locomotion neural components are well known. Both of these central pattern generators (CPGs) have been studied extensively. Despite complete wiring maps of these networks, how they produce their known rhythms remained unknown until possibilities were suggested in the FFF papers [9] and [10].

#### 3.1.6.1. Lobster STG from four oscillations with distinct phases

The STG controls the chewing action in the stomach and the peristaltic movement of food through the pylorus to the gut. The American lobster's pyloric CPG has four oscillations with four distinct phases. A novel oscillator designed only to produce four oscillations produces the phases, which are uniformly distributed, and approximate duty cycles (ratio of high signal duration to period) of the pyloric rhythms in the American lobster (*Homarus americanus*) [9]. The oscillator also shares several other features with the lobster's oscillators. This type of oscillator is apparently new to engineering and fills the major gaps in the well-known family of ring oscillators.

#### 3.1.6.2. Lamprey locomotion by inspection and converting the network to engineering form

The lamprey's undulatory swimming motion is controlled by a simple network of neurons in each body segment. Inspection of the known network diagram and converting it to engineering form led to three surprises that were apparently unknown previously. First, the network is composed mostly of neurons functioning as AND-NOT gates (which are virtually never used in engineering) [10]. This is solid evidence that nervous systems use neuron AND-NOT gates. Second, the swimming motion is controlled by out-of-phase signals to the muscles on opposite sides of the body [10]. Third, the segmental component is the most popular flip-flop design in electronic computational systems, with modifications that cause all of the neurons' states to oscillate [10]. This is apparently the first discovery that a known network of neurons functions as a logic circuit. The lamprey's oscillator design may be new to engineering.

The lesson from this discovery is that a known network of neurons should be redrawn in engineering form to understand it.

### 3.2. DEEP Neural Networks

The FFF method can result in a neuronal network that is dynamic, explicit, evolutionary, and predictive (DEEP). These four properties are necessary for a network to show how neurons are actually connected to process information with the speed of most brain functions. The FFF papers apparently contain the only neuronal networks in the literature that have these properties.

The networks' **dynamic** operation means the only changes are the strengths of the neurons' signals, depending on strength changes in the neurons' inputs. No network change is required, such as neurogenesis, synaptogenesis, or pruning, nor is any change required in the way neurons function, such as a change in synaptic strength or the strength of action potentials. This makes the networks' speed consistent with the speed of most brain functions.

All neurons, connections, and types of synapses are shown **explicitly**, and all assumptions of neuron capabilities are stated explicitly. Only minimal neuron capabilities are assumed, and no network capabilities are assumed.

The networks are **evolutionary** in the sense that they provide a biologically advantageous function. This includes networks that produce phenomena whose functions are uncertain, such as the matched periods of neural activity found in EEGs.

Finally, the networks are **predictive** of nervous system phenomena. That is, based on the explicit connections and neuron capabilities, it can be demonstrated that the models generate known nervous system phenomena, and they may make testable predictions of phenomena that are as yet unknown.

### 3.3 Explanatory capabilities of DEEP Neural Networks

Neuroscience philosopher C.F. Craver described different types of models and what is required of a model to be explanatory [17]. The FFF models are extraordinarily explanatory according to his description, especially in the many known phenomena they produce that are by-products of the models' designs and unrelated to the functional properties for which the model was designed.

> "Models are explanatory when they describe mechanisms."

The FFF models provide explicit mechanisms showing how neurons can be connected to generate the major known phenomena of several brain functions.

> "[T]he components described in the model should correspond to components in the mechanism..."

Both the FFF models and nervous systems are composed of neurons.

> "*How-actually* models describe real components, activities and organizational features of the mechanism that in fact produces the phenomenon."

The FFD models' results depend only on explicitly shown excitatory and inhibitory connections and known, minimal neuron capabilities of excitation and inhibition.

> The model should produce "…byproducts or side-effects of the phenomenon…Byproducts include a range of possible features that are of no functional significance for the phenomenon…but are nonetheless crucial for distinguishing mechanisms that otherwise account equally well for the phenomenon."

This property is especially important for the FFF models. Other than the single function for which each model was designed, all of the phenomena produced by the model are by-products that have no significance in the function that the model was designed to produce.

For example, the fuzzy decoder was designed only to identify the ordering of its input strengths. Yet the model accurately produces major phenomena central to color vision, olfaction, and the physiology and anatomy of the cerebral cortex.

For another example, the cascaded oscillators were designed only to synchronize neural state changes for different functions that require a wide variety of processing speeds, including some functions that require processing as fast as possible. Yet the model accurately produces the known distribution of brain wave frequencies in bands that have an octave relationship, with the correct endpoints and peak frequencies for each band. (Recall that "distribution" does not mean "range.") This complex distribution is entirely determined by an explicit function of the mean and variance of neuron delay times. All of these features are unrelated to the properties for which the oscillators were designed.

## 4. Forms that follow function

A single-neuron logic gate [4] and several networks that were discovered with the FFF method [3-10] are discussed here. Each neuron in the figures may represent several neurons.

### 4.1. The neuron AND-NOT gate from a need for a logic primitive

Processing information is obviously a biologically advantageous function. This can be accomplished by logic circuits composed of logic gates.

#### 4.1.1. The neuron AND-NOT gate

A neuron with the capabilities of excitation and inhibition can function as a logic primitive. In simplest terms, a neuron is active when it has excitatory input *and* does *not* have inhibitory input. This is a logical statement, which means a neuron with excitatory and inhibitory input strengths X and Y can function as a logic gate with an output signal strength that has the logic truth value X AND NOT Y. More precisely, this means that for binary inputs (high or low), the output is high if X is high and Y is low, and otherwise the output is low.

This simple observation of the neuron's AND-NOT capability was apparently first noted in [3] and [4]. The observation was also apparently the first suggestion that inhibition has a role in logic operations. Although the neuron's AND-NOT logic capability may seem obvious, it may not have been noticed earlier (and is still largely unrecognized) because the AND-NOT logic primitive is virtually never used in electronic logic circuit design for technical reasons that have nothing to do with nervous system operations. The AND-NOT function (X AND NOT Y) is not to be confused with the often-used NAND function, which stands for NOT AND, i.e., NOT (X AND Y).

With access to a high input, i.e., logic value TRUE, the AND-NOT logic primitive is "functionally complete" for implementing logic operations. The brain has many spontaneously active neurons that can supply the high input. The neuron's functional completeness was apparently first shown in [4]. Functional completeness is a rather comprehensive logic property. It implies, for example, that sufficiently many neurons functioning as AND-NOT gates can implement everything that can possibly be done in electronic logic circuit hardware and software. Neurons are varied and complex, and neurons clearly evolved with additional capabilities besides excitation and inhibition. But excitation and inhibition by themselves are sufficient for processing information.

A symbol for the AND-NOT logic function is shown in Fig 1.

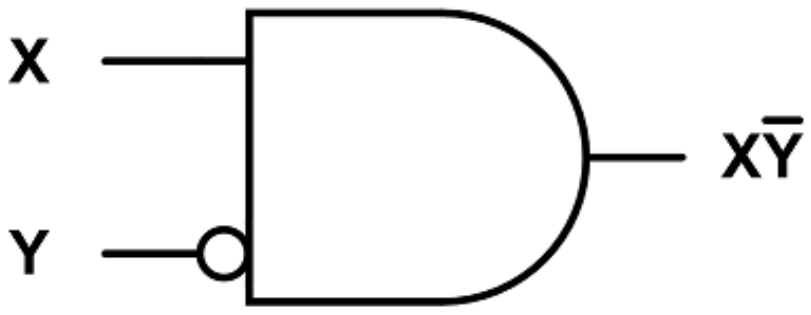

**Fig 1. A standard symbol for the logic primitive X AND NOT Y [7].** The rounded rectangle represents the logic AND function, and the circle represents negation (logic NOT). The AND-NOT gate can be implemented by a neuron with an excitatory input of strength X and an inhibitory input of strength Y. That is, if X and Y have binary strengths (high or low), the strength of the output signal has the Boolean logic truth value X AND NOT Y.

For non-binary inputs, i.e., inputs with strengths that are between high and low, a neuron's output is approximately the "truncated" difference between X and Y: X - Y if $X > Y$ and 0 if $X \leq Y$. The neuron's output strength is truncated because a large inhibitory input generally suppresses a smaller excitatory input. Because of neuron nonlinearities, the neuron's output strength is a measure of the truncated difference rather than the precise difference. The measure of the difference means that if $X > Y$, the response is an increasing function of X, up to some maximum (more excitation increases output), and a decreasing function of Y to some minimum (more inhibition decreases output). For convenience, this measure of the truncated difference will be written as X - Y, and the truncated difference itself will sometimes be used as an approximation of the measure of the truncated difference.

Also for convenience, neuron signal strengths are normalized by dividing by the maximum strength so that the minimum and maximum strengths are 0 and 1 (or 0% and 100%), respectively. For logic functions, the maximum and minimum signal strengths stand for the Boolean logic truth values TRUE and FALSE, and intermediate strengths stand for fuzzy logic truth values. For a neuron signal consisting of spike trains, the strength is measured by the frequency of spikes. This means that for an oscillating signal consisting of periodic bursts of high frequency spike trains, the frequency of the oscillation is the frequency of bursts, not the frequency of spikes.

The AND-NOT logic primitive has simplicity, efficiency, and power that have been underappreciated. For example, AND-NOT gates with the truncated difference response can be connected to find maximum and minimum values of network inputs, as demonstrated in the color vision model below. This cannot be accomplished with arithmetic operations, such as a weighted sum of inputs, or a non-linear function of the sum, that is commonly assumed for neuron responses. Several signaling systems can be modeled with AND-NOT gates because of the availability of excitation and inhibition in diverse components, e.g., a neuron, a transistor, regulatory DNA, an immune system cell, a smoke signal.

There have many proposals for a neuron functioning as a logic gate at least since McCulloch and Pitts, but they all require speculative neuron capabilities or changes in the neuron, such as synaptogenesis, pruning, a change in synaptic strength, or the strength of action potentials. The neuron AND-NOT gate follows from the known neuron capabilities of excitation and inhibition, which can change in real time, i.e., a few milliseconds.

Fig 2 shows an electronic implementation of an AND-NOT gate.

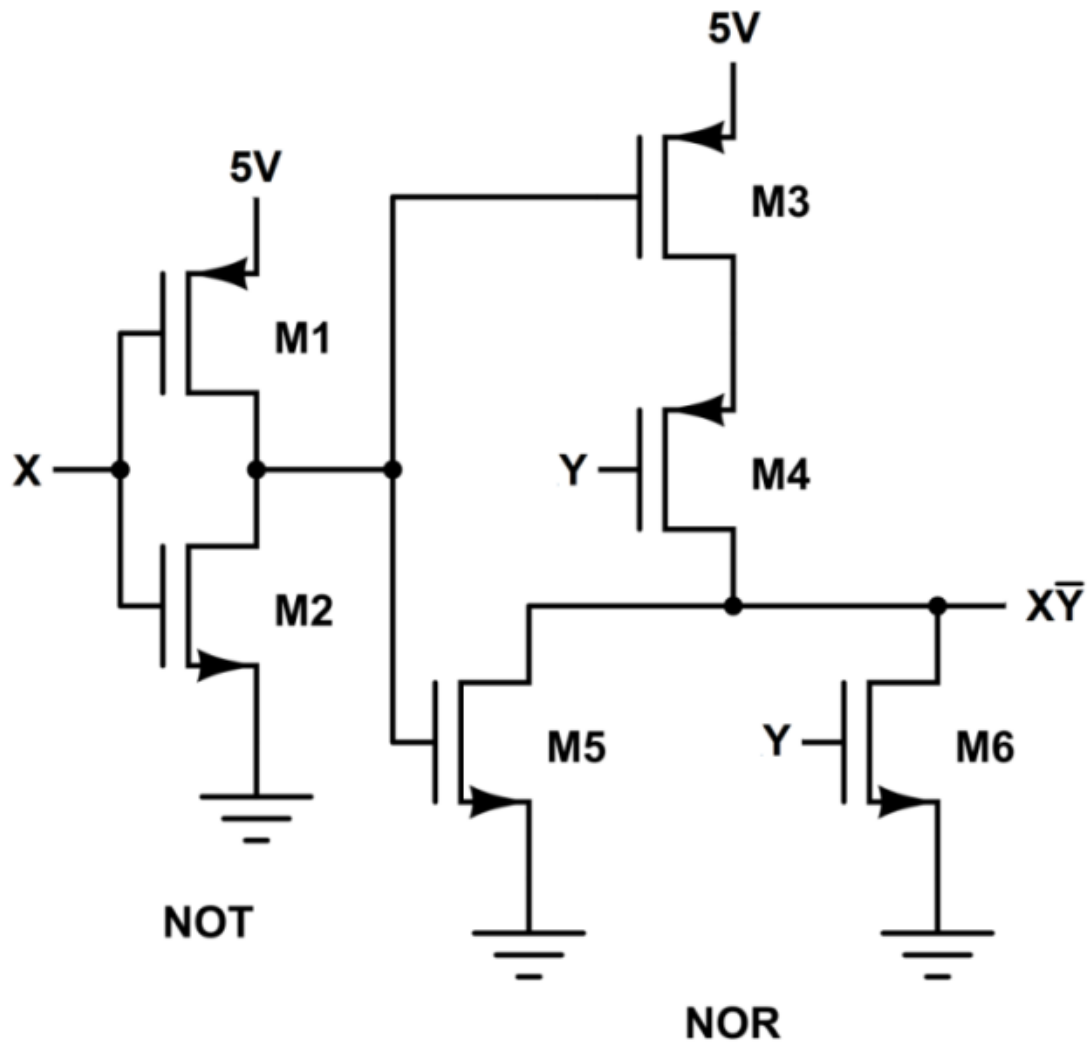

**Fig 2. A Boolean logic AND-NOT gate.** The logic circuit is composed of six transistors, implemented in CMOS architecture. The network consists of a NOT gate followed by a NOR (NOT OR) gate. By De Morgan's law, NOT((NOT X) OR Y) = X AND NOT Y.

Fig 3 shows an electronic implementation of the truncated difference function. This is one of several implementations in [18].

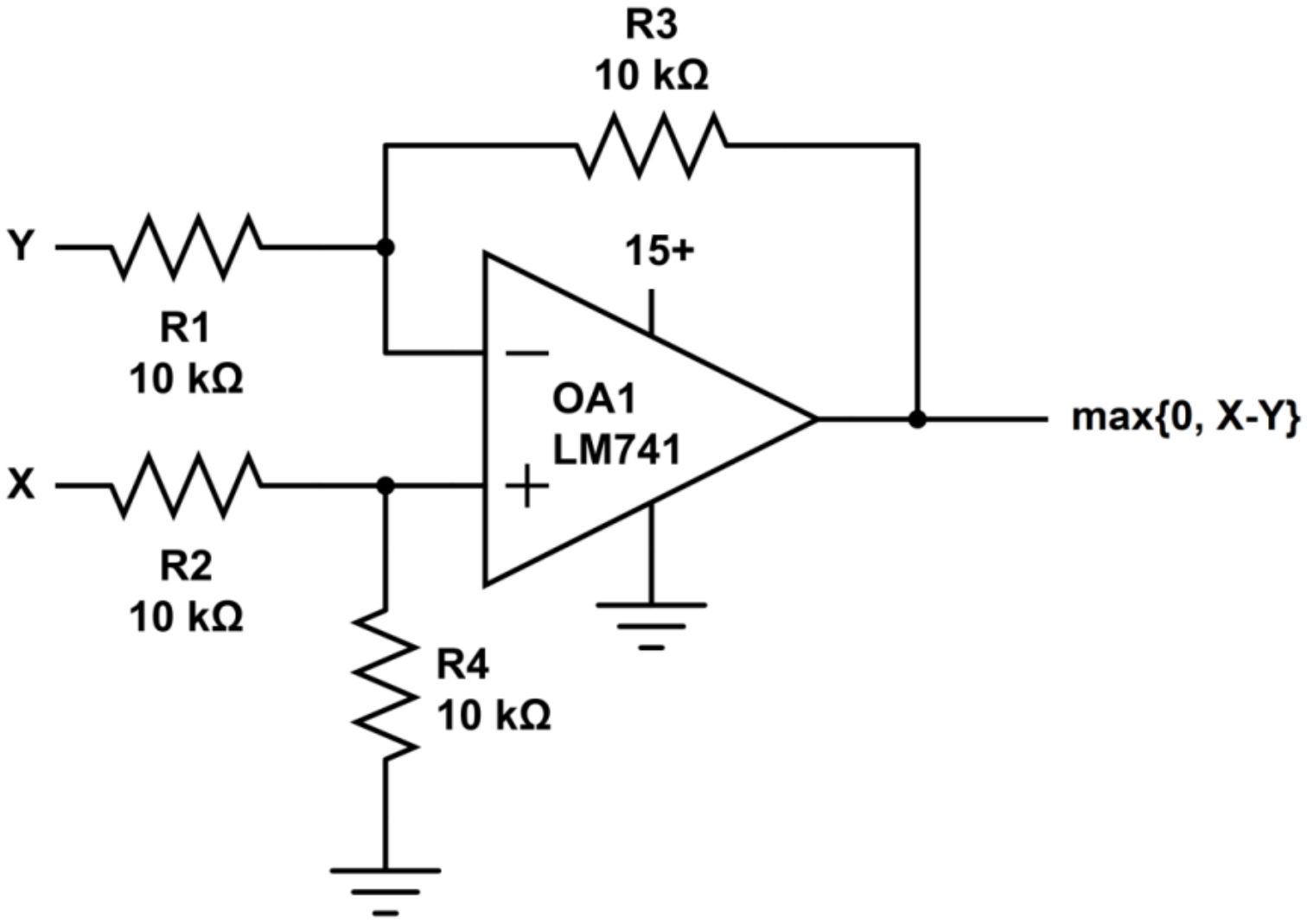

**Fig 3. An implementation of the truncated difference function [18].** The truncated difference of X and Y is max{0, X-Y} (the larger of 0 and X-Y). The truncated difference is an approximation of a neuron's AND-NOT response to an excitatory input of strength X and an inhibitory input of strength Y. The operational amplifier is configured as a subtractor, with the negative differences eliminated by replacing the usual negative voltage power supply with ground.

#### 4.1.2. Noise reduction

For processing digital (Boolean) information, additive noise in binary inputs may affect the intended binary outputs. A sigmoid neuron response reduces moderate levels of additive noise by producing an output that decreases a low input and increases a high input. It may be at least partly for noise reduction that "…the vast majority of neurons show sigmoid nonlinearities" [19].

It will be demonstrated by simulation that a neuron response that is sigmoid in both excitatory and inhibitory inputs is sufficient for the noise-reducing requirements of the flip-flops presented here. But such a response is not necessary; two simpler, more general properties are sufficient.

Reduction of noise in both excitatory and inhibitory inputs can be achieved by a response function of two variables that generalizes a sigmoid function's features. The noise reduction need only be slight for flip-flops because they have feedback loops that continuously reduce the effect of noise.

Let F(X, Y) represent a neuron's response to an excitatory input X and an inhibitory input Y. The function must be bounded by 0 and 1, the minimum and maximum possible neuron responses, and must satisfy the AND-NOT values for binary inputs: F(1, 0) = 1 and F(X, Y) = 0 for the other binary inputs. For other points (X, Y) in the unit square, suppose F satisfies:

1. F(X, Y) > X - Y for inputs (X, Y) in a neighborhood of (1, 0) and

2. F(X, Y) < X - Y or F(X, Y) = 0 for inputs (X, Y) in a neighborhood of the other three vertices of the unit square.

Neurons that make up the networks proposed here are assumed to have these minimal noise-reducing properties.

#### 4.1.2.1. Sigmoid neuron response function used in the simulations

For any sigmoid function f from f(0) = 0 to f(1) = 1, the following function has the noise-reducing properties 1 and 2:

F(X, Y) = f(X) - f(Y), bounded below by 0.

The function F is illustrated in Fig 4 for a specific sigmoid function f. The sine function of Fig 4A was chosen for f rather than any of the more common examples of sigmoid functions to demonstrate by simulation that a highly nonlinear function is not necessary for robust maintenance of binary signals. Fig 4B shows that F is approximately the truncated difference X - Y.

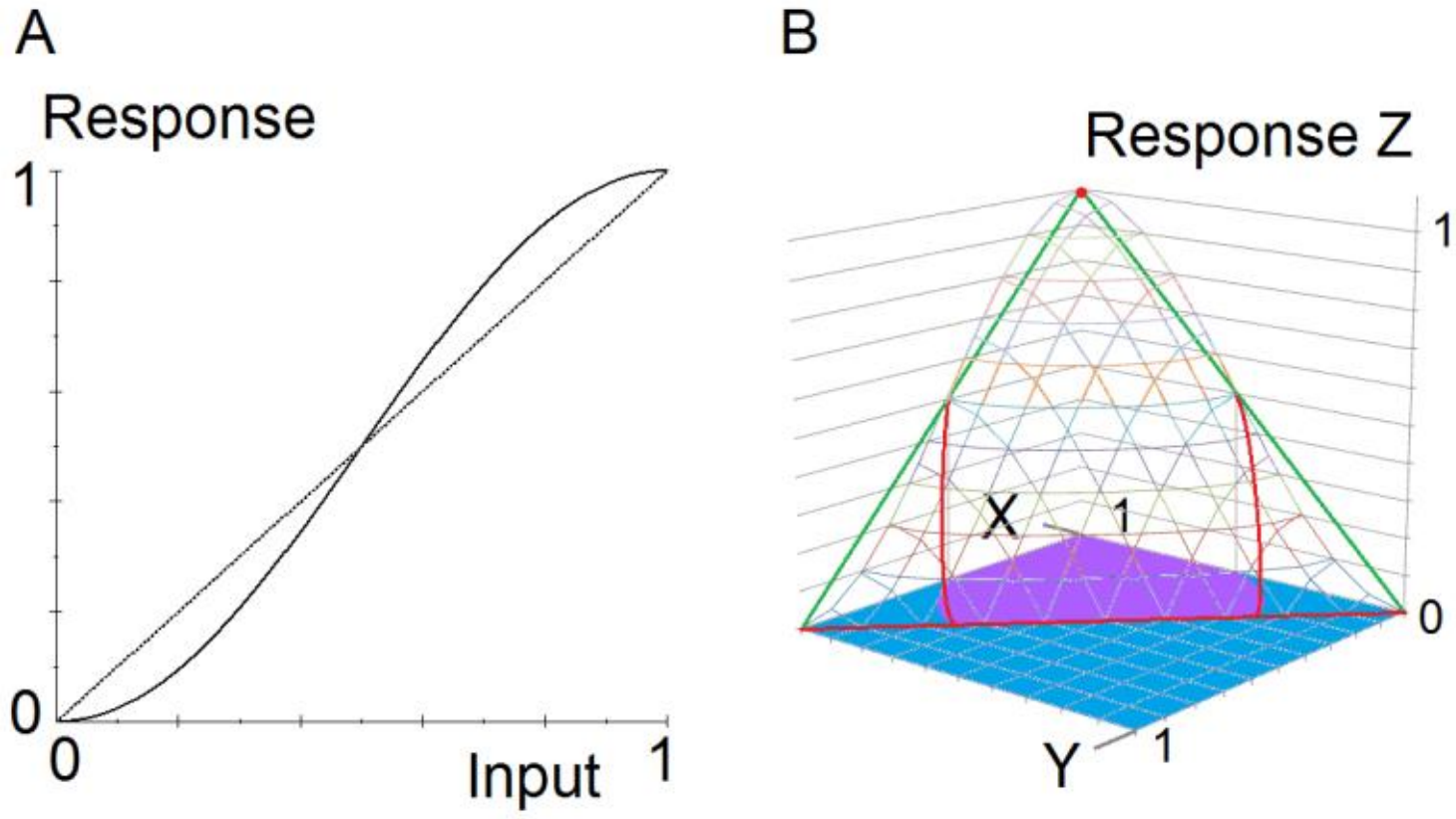

**Fig 4. Noise-reducing AND-NOT function [7].** The graphs show an example of a neuron response to analog inputs that reduces moderate levels of additive noise in binary inputs. **A.** A sigmoid function $f(x) = (1/2)\sin(\pi(x - 1/2)) + 1/2$. **B.** Graph of a function that has the noise-reducing properties 1 and 2. The function is $F(X, Y) = f(X) - f(Y)$, bounded by 0. Wireframe: Graph of the response function $Z = F(X, Y)$. Green and red: A triangle in the plane $Z = X - Y$. Red: Approximate intersection of the plane and the graph of F. Purple: Approximate region in the unit square where $F(X, Y) > X - Y$ (condition 1). Blue: Approximate region in the unit square where $F(X, Y) < X - Y$ or $F(X, Y) = 0$ (condition 2).

**4.1.2.2. A primitive noise-reducing AND-NOT gate**

A response that satisfies conditions 1 and 2 in section 4.1.2 does not indicate capabilities of sophisticated logic or mathematics. An AND-NOT response with properties 1 and 2 can be produced by mechanisms that are quite simple. Fig 5 shows that a single transistor and three resistors can be configured to accomplish this. The network output was simulated in CircuitLab, and the graph was created in MS Excel and MS Paint. The inputs X and Y vary from 0V to 5V in steps of 0.05V. A 5V signal commonly stands for logic value 1, and ground stands for logic value 0.

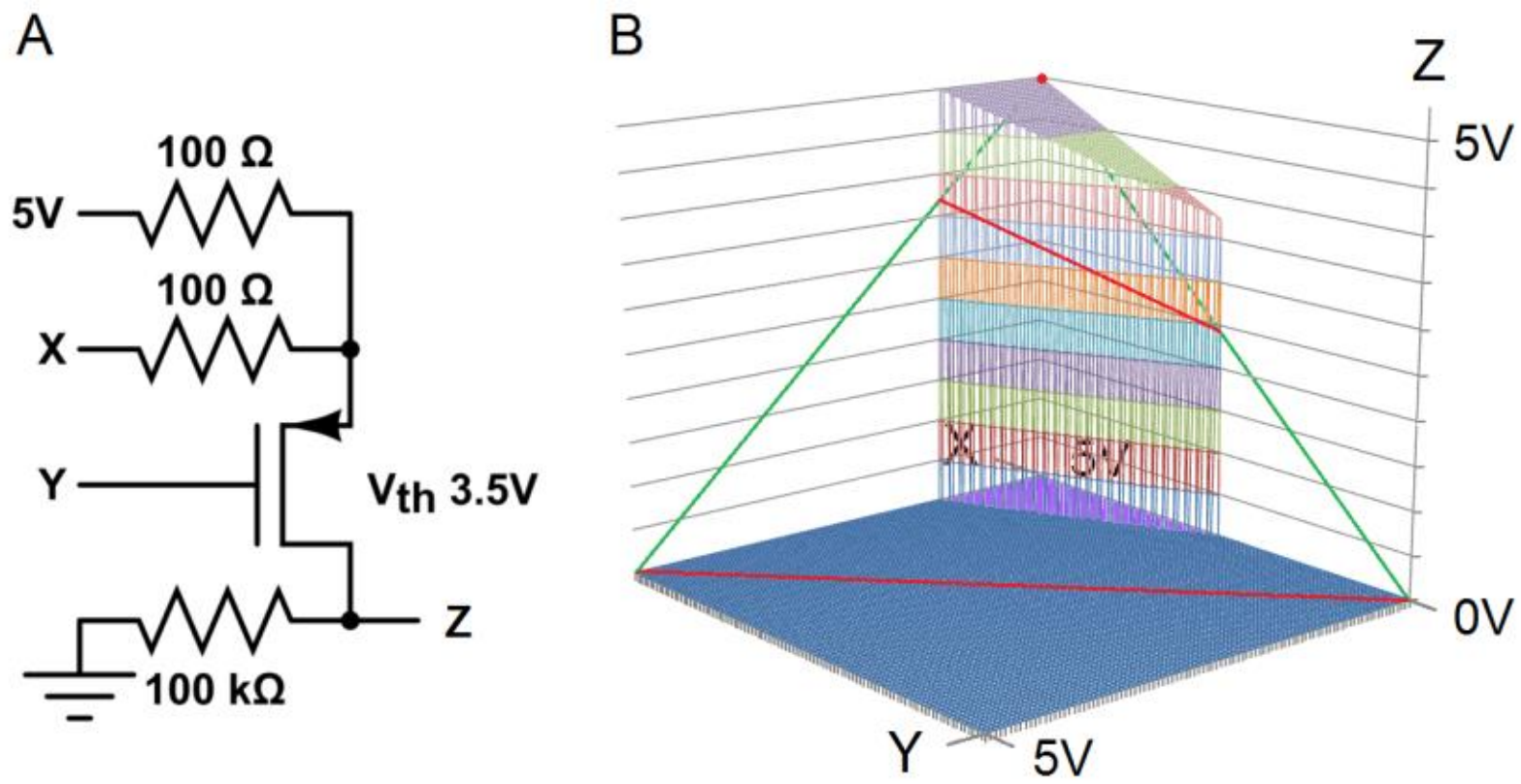

**Fig 5. Single transistor AND-NOT gate that reduces noise [7].** This minimal logic circuit satisfies the noise-reducing conditions 1 and 2. **A.** A logic circuit consisting of one transistor and three resistors. **B.** Engineering software simulation. Wireframe: Graph of the transistor response function Z = F(X, Y). Green and red: A triangle in the plane Z = X - Y. Red: Intersection of the plane and the graph of F. Purple: Region in the unit square where F(X, Y) > X - Y (condition 1). Blue: Region in the unit square where F(X, Y) < X - Y or F(X, Y) = 0 (condition 2).

### 4.2. Color vision from consistent discrimination of spectral distributions under different lighting intensities

Distinguishing spectral distributions in different photostimuli is an obviously advantageous function for distinguishing objects. This can be accomplished with different classes of photoreceptors that are more sensitive to different regions of the visible spectrum. As stated above, the problems are that all photoreceptors are somewhat sensitive to all parts of the spectrum, and the intensity of natural light varies throughout the day and when the weather or shelter changes. The change in light intensity changes the intensity of photostimuli, which in turn changes the response strengths of color photoreceptors (cones). The responses of different classes of cones change by different amounts, depending on how close each one is to saturation or the resting level. But the ordering of the response strengths remains constant. Humans have three classes of cones. Identifying the ordering of the three response strengths is the key to identifying the spectral distribution under varying lighting intensities. This capability is usually called color constancy, but a different color perception for each ordering turns out to be the main function of color vision.

### 4.2.1. A fuzzy decoder for identifying ordering of input strengths

In digital electronics, a binary decoder is a logic circuit that has n binary inputs and $2^n$ outputs, one for each possible combination of inputs. Each input has one of two possible values, usually called 0 or 1, so there are $2^n$ possible combinations of inputs. The outputs identify the combination of inputs. For example, a decoder with two inputs has four possible combinations of input values, 00, 01, 10, and 11. If the input combination is 01, the output for identifying that combination is 1 and the other outputs are 0. If 1 and 0 stand for logic truth values TRUE AND FALSE, the output can also stand for the logic truth value of the statement "The first input is FALSE and the second input is TRUE."

Color vision is a decoder. The three classes of cones encode the spectral information in the photostimulus. Color vison decodes the encoded information. For example, if S, M, and L stand for Boolean response strengths of each of the cone classes, SML = 110 stands for red, 100 stands for yellow, etc. (Photoreceptors are spontaneously active, and light suppresses their response strengths.)

That color vision is a decoder does not depend on any particular model. It is another simple observation that was apparently not previously recognized, possibly because the color vision decoder uses fuzzy, or analog, logic and electronic logic circuits generally use Boolean, or digital, logic.

#### 4.2.1.1. The fuzzy decoder

Photoreceptors have graded responses, i.e., response strengths between 0 and 1. Identifying the ordering of photoreceptor response strengths requires an analog, or fuzzy logic, decoder, i.e., logic with truth values of intermediate strengths. A retinal fuzzy logic decoder, proposed in [3] and refined in [4], performs the color constancy function of identifying the ordering of the three photoreceptor response strengths. The model is illustrated in Fig 6.

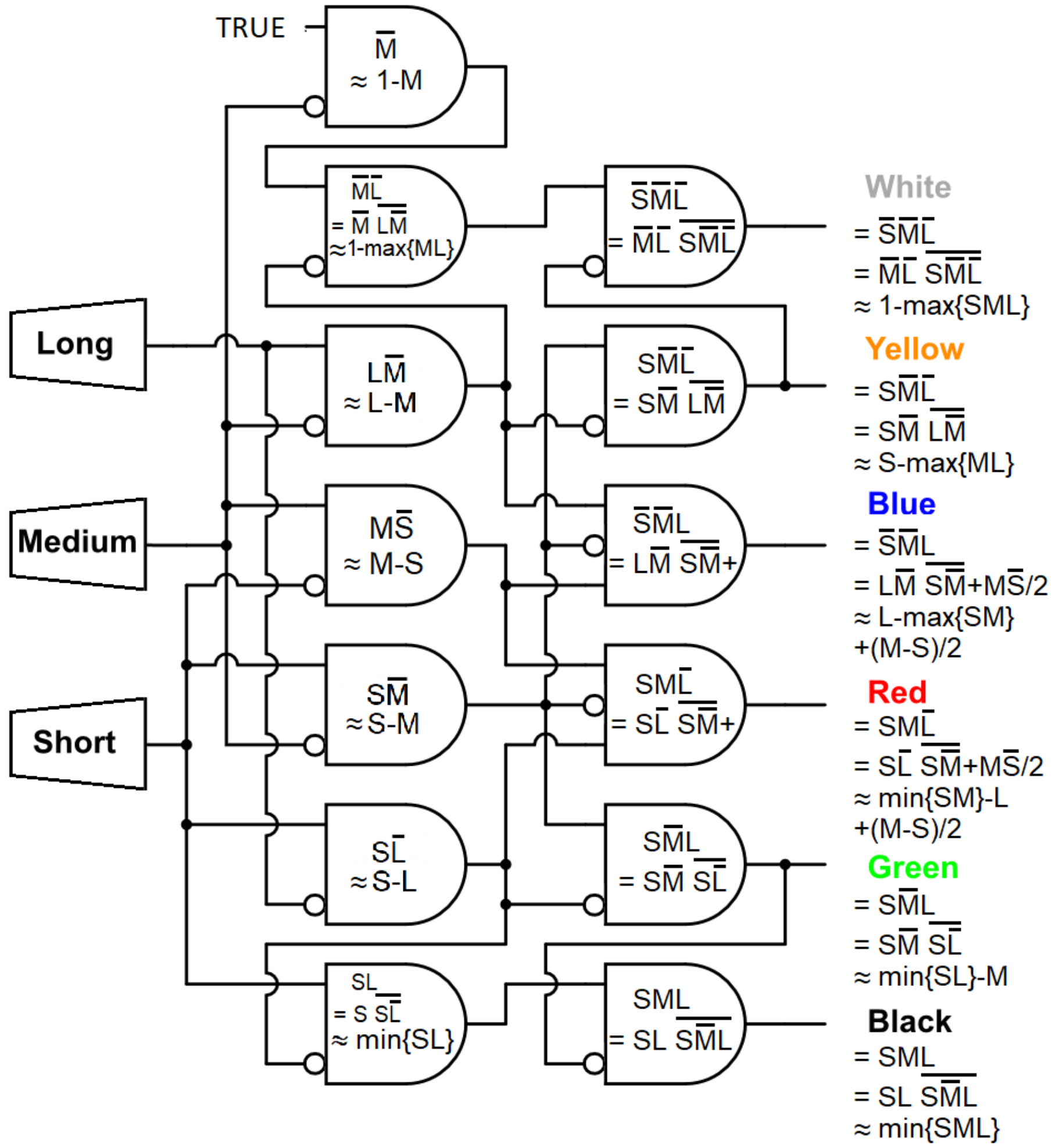

**Fig 6. Fuzzy decoder model of color vision.** The network receives input from three classes of spatially proximate photoreceptors that are sensitive to short, medium, and long wavelengths. The six outputs are neural correlates of the perceptions of color vision, including black and white. All cones are assumed to receive input from the same photostimulus.

Psychophysical tests have shown we can perceive four colors – red, green, blue, and yellow – and black and white. The color vision model has six outputs instead of the complete decoder's eight because human color vision conveys violet and purple information through the red and blue channels rather than dedicated channels. In a complete decoder, the neural correlate of the strength of the perception of purple is M-max{SL}. For violet, the strength is min{ML}-S. So in any case, the total strength of violet and purple is M-S. (These differences are truncated at 0.) Accordingly, M-S is distributed equally to the Red and Blue cells in Fig 6. More is said about violet and purple in section 4.2.2.4.3 and Fig 10 in section 4.4.2.

Fig 7 shows the fuzzy decoder model of Fig 6 as depicted in three two-minute YouTube videos that illustrate the main features of the decoder. Links to the YouTube videos are on my website at NeuralNanoNetworks.com > Color Vision > Videos. The simplified implementation of the color vision model in the videos uses the truncated difference of the inputs for the approximation of each neuron's output. Fig 7 alone illustrates several of the decoder's main features.

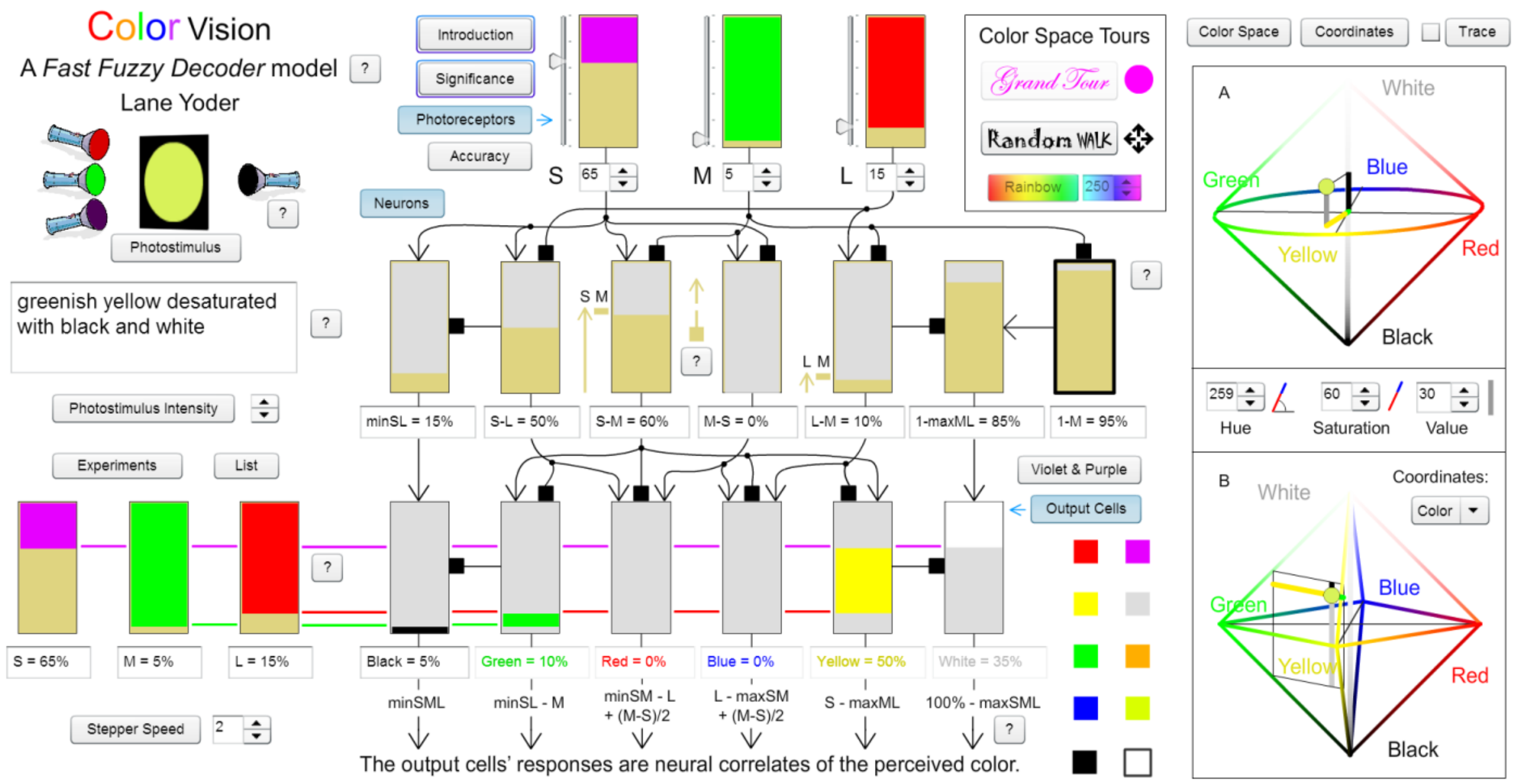

**Fig 7. A different representation of the network in Fig 6.** The cones at the top are redrawn at the bottom left to show the relation of the network inputs to the outputs. Arrows indicate excitatory input; blocks indicate inhibitory input.

The decoder's outputs are measures of the intervals defined by the three inputs. The color cells that respond identify the ordering of the inputs. For example, Fig 7 shows a photostimulus that produces cone responses M < L < S. For this ordering, the green and yellow cells respond. A different ordering would produce responses from a different combination of color cells. This is illustrated in the YouTube videos. Since all inputs from the cones are between (not equal to) 0 and 1, the black and white cells also respond.

Each output cell's response strength is the neural correlate of the strength of perception. Together, the responses in Fig 7 are the perception of greenish yellow, desaturated with black and white. At least two of the color cells' fuzzy truth values is always 0. The outputs' total truth value is always (approximately) 1, the same as a Boolean decoder with one output 1 and the rest 0.

#### 4.2.1.2. Defining a network connectivity

Boolean logic identities can define network connectivity. For example, in Boolean logic, yellow is perceived when SML = 100. That is, Yellow = S AND NOT M AND NOT L. The most obvious implementation with two AND-NOT gates is (S AND NOT M) AND NOT L, or (S-M)-L. For any logic statement, there are infinitely many fuzzy logic functions that agree with the Boolean values. Not all functions can identify the ordering of the inputs. The fuzzy logic produced by the obvious implementation above does not identify the cone response ordering. A correct implementation to identify the ordering is
(S AND NOT M) AND NOT (L AND NOT M), or (S-M)-(L-M). This implementation of Yellow is shown in Figs 6 and 7.

### 4.2.2. Color phenomena produced by the fuzzy decoder model

#### 4.2.2.1. A partial list

In the interest of brevity, a few phenomena that were shown in [3] to be produced by the fuzzy decoder model are only mentioned in this subsection. The fuzzy decoder model produces color mixing that results in an entirely different color (illustrated for yellow from red and green in the first YouTube video link on the website), opponent-color cells, chromatic color distinctions, the achromatic characteristics of black and white, and the continuous yet categorical nature of color (best illustrated in a rainbow).

#### 4.2.2.2. Color space

The fuzzy decoder model's geometrical color space provides detailed, quantitative explanations of standard color figures such as Newton's color wheel and Munsell's color system. The geometry of the rectangular and cylindrical coordinate systems is outlined here and explained in more detail in [3] and at
NeuralNanoNetworks.com > Color Vision > Color Coordinates.

##### 4.2.2.2.1. Rectangular coordinates

The decoder's six outputs are rectangular coordinates of perceived color space, i.e., the way subjects arrange colored chips closely together if they appear to be nearly the same color. This geometrical figure is illustrated in the lower right of Fig 7 (Figure 2 in [3]). The figure can be visualized because two of the color coordinates are zero and their sum is (approximately) one. So the six coordinates define four three-dimensional tetrahedrons that are joined at their common faces and embedded in six-dimensional space.

#### 4.2.2.2.2. Cylindrical coordinates

Simple functions of the fuzzy decoder's six rectangular coordinates also provide cylindrical coordinates for the color space as hue, saturation, and value [3]. The cylindrical coordinates are not part of the fuzzy decoder model. This more common depiction of color space is illustrated in the upper right of Fig 7 (Figure 3 in [3]). The two geometrical figures are topologically equivalent. They represent the same color space of fuzzy decoder responses but with different coordinate systems. The short explanation is that horizontal squares in rectangular coordinates are transformed into circles in cylindrical coordinates because the radius coordinate is constant on the squares.

#### 4.2.2.3. Unique spectral colors

For a spectral photostimulus (single wavelength or narrow band), unique blue, green, and yellow are perceived at, or near, 470, 510, and 575 nm, respectively [20-22]. (These values can be determined within 1 or 2 nm for individuals, but there is some inter-observer variability.)

The decoder model's property of producing a different combination of color cell responses for different orderings of cone responses implies that a single color cell responds at each of the three intersections of the cones' sensitivity curves. This decoder property is consistent with a visual comparison of the curves and the three unique color wavelengths.

The close fit of unique color wavelengths and sensitivity curve intersections may not have been reported in the literature because researchers are evidently more interested in curve peaks than intersections. It may not have been noticed that while the common practice of normalizing the sensitivity curves to have the same heights does not affect the peak wavelengths, it does change the intersections.

#### 4.2.2.4. Mutually exclusive color pairs

The search for an explanation of mutually exclusive colors has a 200-year history with heated debates that went to the core of how the brain processes spectral information. Two of the six possible color pairs cannot be perceived together. No color is described as reddish green or bluish yellow.

The fuzzy decoder model has a simple explanation based on the model's property of producing a different combination of color cell responses for different orderings of cone activity. The model implies that red is perceived when the L cones absorb the most photons; green is perceived when M cones absorb the most. They cannot both absorb the most. The fuzzy decoder model implies that yellow is perceived when S absorbs the least; blue when L absorbs the least. They cannot both absorb the least.

That these decoder model implications agree with perception can easily be seen for spectral colors, at least approximately, by comparing the visible spectrum, e.g., a rainbow, with the S, M, and L sensitivity curves.

#### 4.2.2.5. Perceptual anomalies

The fuzzy decoder model produces some of the same "shortcomings" as the brain's color perceptions.

##### 4.2.2.5.1. Bezold-Brücke hue shift

Color perception is not quite constant under different photostimulus intensities. Both orange and greenish-yellow appear yellower at higher intensities. Violet and greenish-blue appear bluer. This phenomenon is the Bezold-Brücke hue shift [23, 24]. The fuzzy decoder model produces this result, with the same colors [3].

#### 4.2.2.5.2. Additivity failure of brightness

For similar reasons, the fuzzy decoder model produces what is known as the "additivity failure" of brightness: When two photostimuli are superimposed, the perceived brightness is not additive if the stimuli have sufficiently different spectral distributions [25-27]. The effect is not due to nonlinearities in neural responses.

#### 4.2.2.5.3. Violet and purple

Psychophysical evidence shows that violet and purple information is transmitted through the red and blue channels. This may be because the benefit of complete violet and purple information transmitted through dedicated channels was outweighed by the cost of increasing the size of the optic nerve by a third and increasing the brain correspondingly to process the information. In the fuzzy decoder model, the red and blue channels are the only channels that can convey the violet and purple information without a considerable loss of information [3].

### 4.2.2.6. Center-surround

Center-surround phenomena may be the best-known feature of retinal processing of spectral information. It was recently shown that the color vision model discussed here (Fig 6) can produce these phenomena [6]. Apparently this is the first explanation of the center-surround phenomena. The model includes both single- and double-opponent cells and implies that center-surround and mutually exclusive colors are two different kinds of color opponency.

## 4.3. Olfaction from consistent discrimination of odorants under different concentrations

Like color constancy for different photostimuli intensities, olfaction needs to identify odorants independently of their concentrations. Humans have approximately 300 classes of olfactory sensory cells (estimates vary). The set of recursive logic identities in Table 1 (from [4]) extended the fuzzy decoder in [3] to allow any number of inputs. Like the color vision decoder, the combination of positive outputs identifies the ordering of the inputs. With inputs from olfactory sensory cells, the decoder can identify odorants independently of their concentrations. This decoder was also shown in [4] to produce olfaction’s four main properties, which are analogous to properties of color vision. (A decoder with n inputs has $2^n$ outputs, so clearly human olfaction could not have a complete decoder, nor would it be likely to have a use for so much information. Few of the output cells would ever have a positive response.)

| Recursive and reductive logic identities for a fuzzy decoder composed of AND-NOT gates | Interval measured by the decoder response | Approximate value of the decoder response |
|---|---|---|
| 1. $\Pi_{i=1}^{M} X_i \Pi_{j=1}^{N} \overline{Y_j} = \{\Pi_{i=1}^{M-1} X_i \Pi_{j=1}^{N} \overline{Y_j}\}\overline{\{\overline{X_M} \Pi_{i=1}^{M-1} X_i \Pi_{j=1}^{N-1} \overline{Y_j}\}}$ | $[\max_{j=1}^{N}\{Y_j\},\ \min_{i=1}^{M}\{X_i\}]$ | $\min_{i=1}^{M}\{X_i\} - \max_{j=1}^{N}\{Y_j\}$ |
| 2. $\Pi_{i=1}^{M} X_i \Pi_{j=1}^{N} \overline{Y_j} = \{\Pi_{i=1}^{M} X_i \Pi_{j=1}^{N-1} \overline{Y_j}\}\overline{\{Y_N \Pi_{i=1}^{M-1} X_i \Pi_{j=1}^{N-1} \overline{Y_j}\}}$ | | |
| 3. $\Pi_{i=1}^{N} X_i = \{\Pi_{i=1}^{N-1} X_i\}\overline{\{\overline{X_N} \Pi_{i=1}^{N-1} X_i\}}$ | $[0,\ \min_{i=1}^{N}\{X_i\}]$ | $\min_{i=1}^{N}\{X_i\}$ |
| 4. $\Pi_{j=1}^{N} \overline{Y_j} = \{\Pi_{j=1}^{N-1} \overline{Y_j}\}\overline{\{Y_N \Pi_{j=1}^{N-1} \overline{Y_j}\}}$ | $[\max_{j=1}^{N}\{Y_j\},\ 1]$ | $1 - \max_{j=1}^{N}\{Y_j\}$ |

**Table 1. Logic identities that define a fuzzy decoder.** The recursive and reductive identities in the first column show how fuzzy decoders can be implemented with AND-NOT gates. The logic identities equate every product (conjunction) to a product A AND NOT B, where A and B each have one fewer components than the first product. Equations 1 and 3 show how to add an X to a product, and equations 2 and 4 add a NOT Y. The second column shows the interval measured by the corresponding network response. The third column shows the value of the response is approximately the length of the interval.

### 4.4. Cerebral cortex physiology and anatomy from the modified decoder

The extended decoders exhibit major features of the cerebral cortex’s physiology and anatomy [5]. As mentioned above, these results show how surprisingly powerful the FFF method can be. No advantageous function was considered for this discovery. It was simply observed that the extended decoders exhibit several of the major features of the cerebral cortex. The next subsection gives a highly useful function that provides a plausible explanation for this unexpected result. This explanation was not included in [5].

#### 4.4.1. Modified decoders for all combinational processing of information

With the addition of one neuron, the extended decoders in [4] are general logic circuits that can provide the foundation for all of the brain's combinational processing of information, i.e., all information processing except sequential. A fundamental theorem of logic states that every Boolean function can be expressed as a disjunction of conjunctions. The extended decoder with binary inputs produces all conjunctions of the inputs. A Boolean disjunction (OR logic gate) can be implemented by a single neuron with any number of excitatory inputs and no inhibitory input. This follows from the same argument used for the neuron's AND-NOT capability: A high input from any one of the excitatory inputs can produce a high output. So any Boolean function can be realized with a disjunction of some combination of the outputs of a decoder.

#### 4.4.2. Cerebral cortex physiology and anatomy

The addition of one neuron to a decoder is small compared to the decoder, especially a decoder with many inputs. Nearly all of the physical properties of the modified decoder will be in the decoder itself. If the decoders' total cost is optimized, i.e., if the number of neurons is minimized and the neurons are arranged to minimize total connection length (distance between soma of connected pairs of cells) and maximize packing density, the decoders exhibit major local characteristics of the anatomical structure of the cerebral cortex [5]. The decoders also produce correlates of physiological brain phenomena. These properties are illustrated in Figs 8-14 and are discussed below.

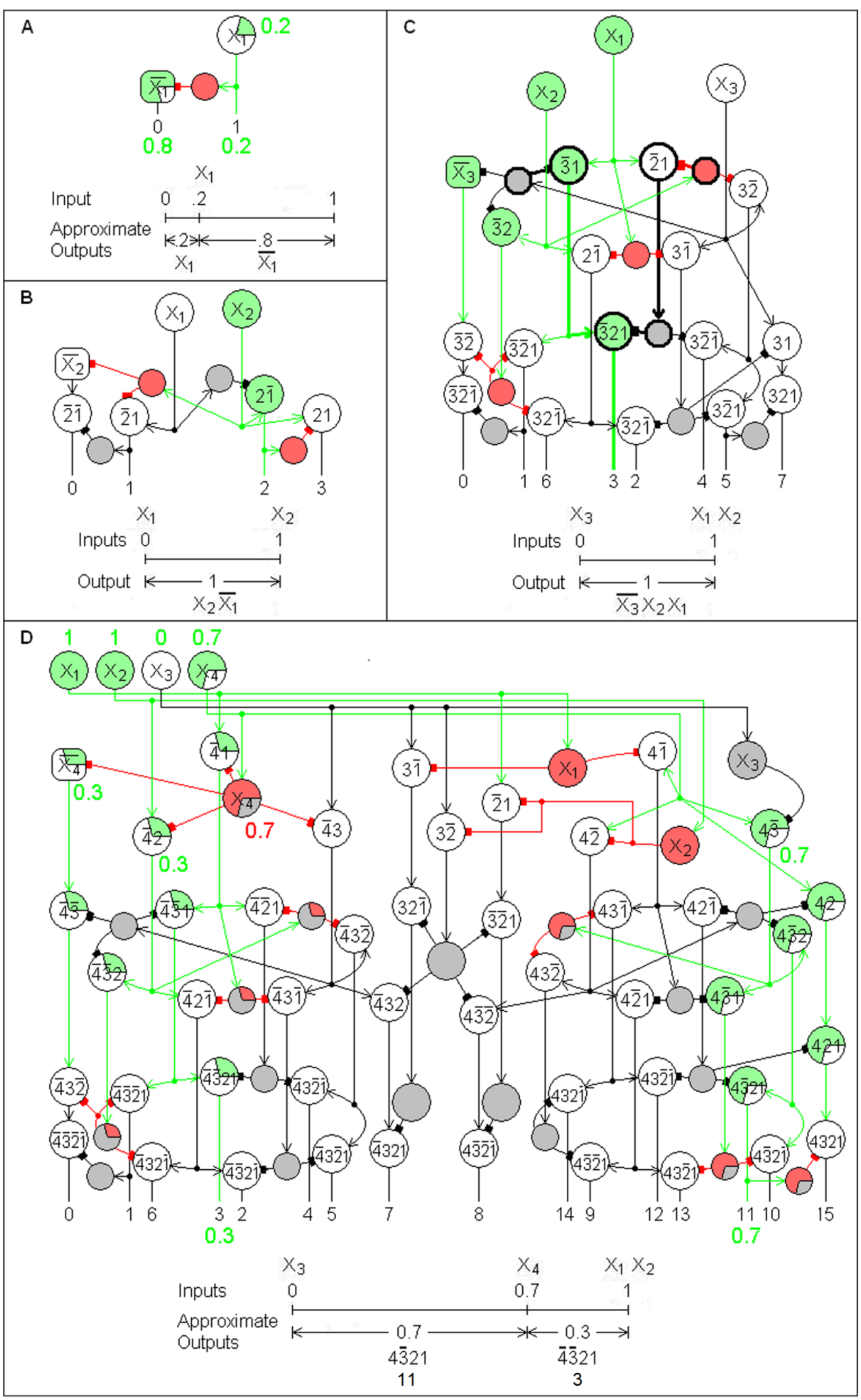

**Fig 8. Fuzzy Decoders [5].** A fuzzy decoder with n inputs produces $2^n$ outputs corresponding to the $2^n$ possible combinations of binary inputs. The figure shows fuzzy decoders with 1 to 4 inputs. Their construction is defined by the logic identities in Table 1, and the cells are arranged to maximize packing density and minimize total connection length. The label on each neuron represents its response, as explained below. Arrows indicate excitatory input; blocks indicate inhibitory input. Spontaneously active neurons are square. To illustrate example inputs and outputs, active neurons are colored. The colored portions and adjacent numbers show the output strengths of the active neurons. Inactive inhibitory cells are shaded. The line graphs below each diagram illustrate the (approximate) interval measure property and the Boolean and fuzzy logic of the example inputs and outputs.

The labels in the circles in Fig 8 are shorthand for the cell's logic output. For example, the cell with output labeled 3 at the bottom of Fig 8D has the fuzzy truth value output NOT $X_4$ AND NOT $X_3$ AND $X_2$ AND $X_1$, i.e., the truth value of the statement X4 and X3 are false and X2 and X1 are true. According to Table 1, the value is approximately $\min\{X_2, X_1\} - \max\{X_4, X_3\} = 0.3$. The output label at the bottom is 3 because the Boolean output is 1 when the Boolean inputs are $X_4X_3X_2X_1 = 0011$, which is the binary expression for the number 3.

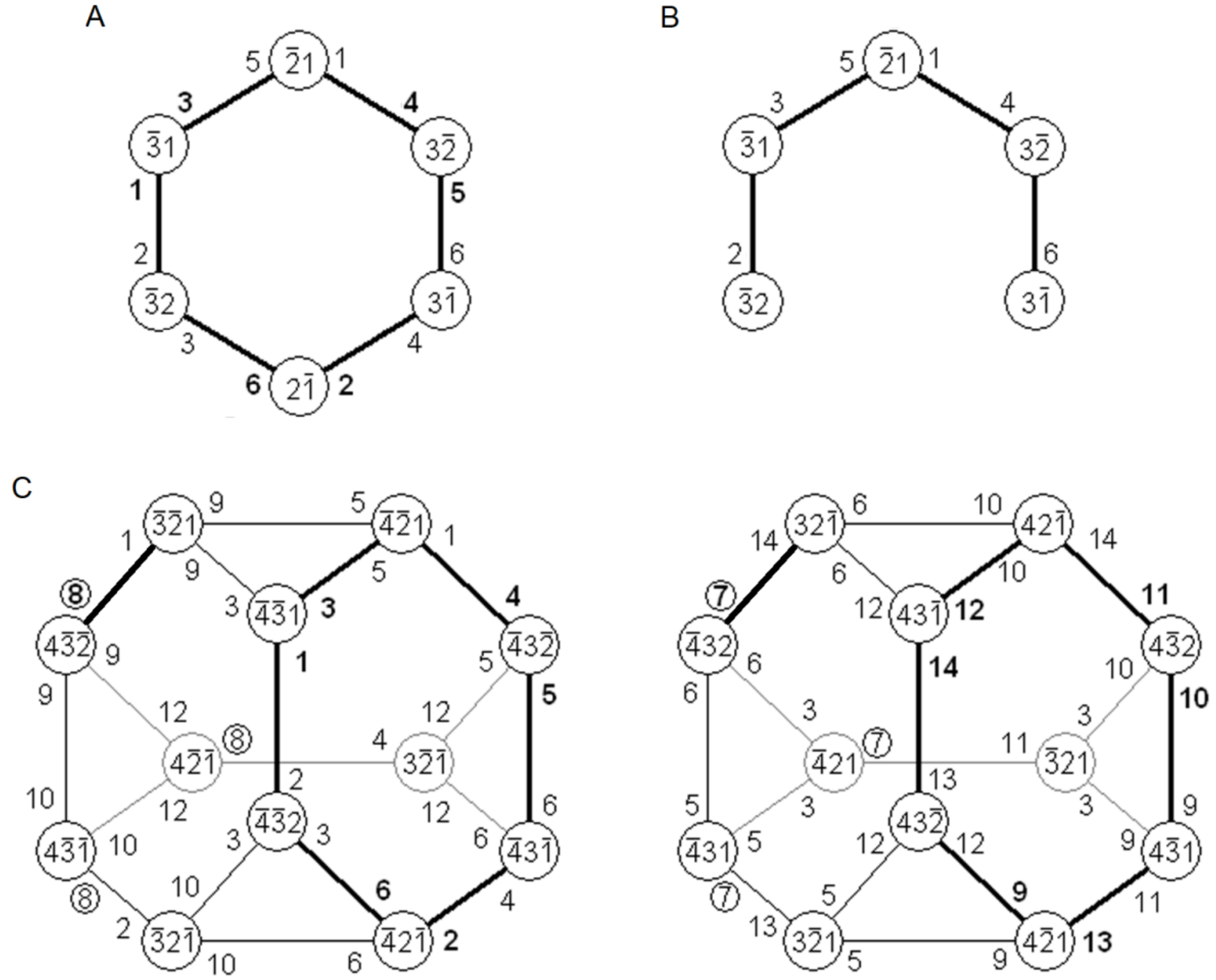

**Fig 9. Fuzzy decoder architecture [5].** The logic identities of Table 1 determine the synaptic connections of fuzzy decoders. For decoders with three and four inputs, the graphs show all of the possible connections according to the logic identities 1 and 2 of Table 1. The nodes are labeled the same as the cells in Fig 8. Two nodes are linked if the cells they represent can be connected by an AND-NOT function according to equation 1 or 2 of Table 1. The numbers next to the nodes represent the network outputs as in Fig 8. Along each edge linking two nodes $N_1$ and $N_2$, the number next to $N_1$ represents the decoder output $N_1$ AND NOT $N_2$. The nodes in graph **A** show all of the ways two of three inputs can be connected with the AND-NOT function. The numbers next to the nodes show that each of the six possible outputs can be produced in two ways. The numbers in boldface show the six outputs as they are implemented in Fig 8C. Graph **B** shows that all six of the outputs can be produced with five nodes. The nodes in graph **C** show all of the ways three of four inputs can be connected according to identities 1 and 2 of Table 1. The numbers in boldface show the outputs numbered 1-14 as they are implemented in Fig 8D.

The graph in Fig 9C forms two, disconnected, truncated tetrahedrons. The disconnection illustrates that for four or more inputs, a decoder must be implemented in separate columns. Two of the outputs must be implemented outside of the columns (7 and 8 in the boldface example and Fig 8D). Also, as illustrated in Fig 8, for n inputs, outputs 0 and $2^n - 1$ must be implemented outside of the columns according to identities 4 and 3 of Table 1, respectively. A decoder with n inputs has $2^n$ outputs, and each column has 8 outputs (including two outside the column). So if n $\geq 3$, $2^{n-3}$ separate columns are required.

Fig 10 illustrates Fig 9A with two possibilities for the color vision decoder to convey violet and purple information through the red and blue channels. A model with separate cells for Violet and Purple was not included in [3] or [4]. This new model may be a more realistic possibility for color vision than the original version in Figs 6 and 7.

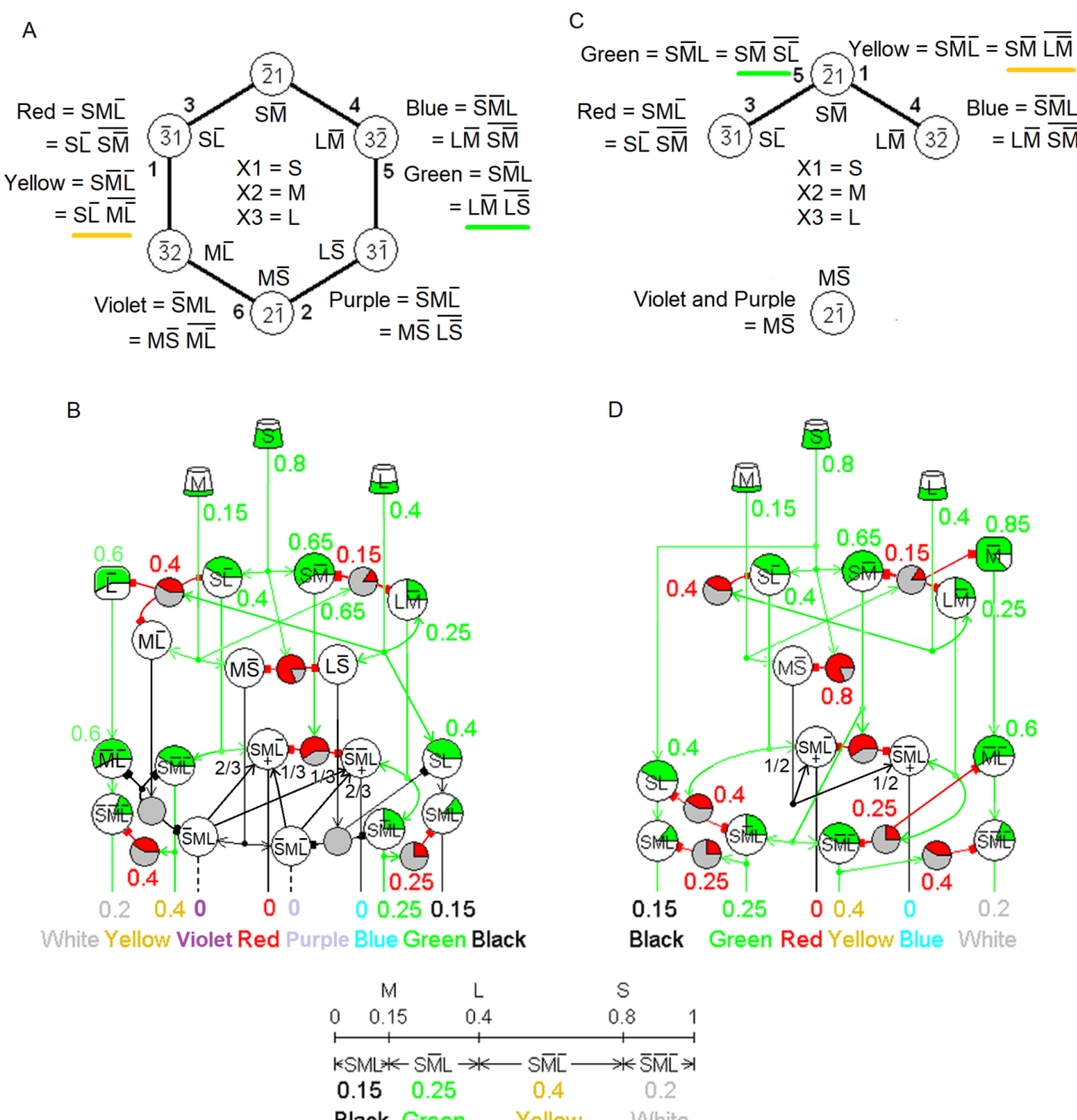

**Fig 10. Possibilities for color vision.** The diagrams show two implementations of the color vision model. Fig A shows all six possible outputs of the graph in Fig 9A with inputs from the three retinal cone classes. The model includes Violet and Purple cells. Fig B shows the model implemented as Fig 8C, as the cells could be arranged optimally in the 3D retina. Fig C shows

part of the graph in Fig 9A representing the minimalist color vision model in Figs 6 and 7. Fig D shows the model arranged in 3D.

Including the Violet and Purple cells allows different input strengths to the Red and Blue cells, as shown in Fig 10B, which could make the perceptions of violet and purple more distinguishable than the equal inputs in Fig 10D. Because of this, Figs 10A and B may be more accurate as retinal implementations for violet and purple than the original model in Figs 6, 7 and 10C and D.

The different ways of connecting the cells for Green and Yellow put these two cells on opposite sides of the columns in Figs 10B and D. This requires different connections for Black and White for efficient connections. The different ways of obtaining these four outputs give approximately equal results, but not because the computations are the same. They are equal because the connections according to Table 1 and the fuzzy logic of the AND-NOT gate output as the truncated difference of the inputs make the outputs of the different networks measures of the same intervals between inputs. But the computations are quite different. For the example inputs in Fig 10, the two different computations for Green are given below. The truncated difference is underlined.

$$\text{Green}_1 \approx (\text{L-M})\text{-}(\text{L-S}) = (0.4\text{-}0.15)\text{-}(\underline{0.4\text{-}0.8}) = 0.25$$

$$\text{Green}_2 \approx (\text{S-M})\text{-}(\text{S-L}) = (0.8\text{-}0.15)\text{-}(0.8\text{-}0.4) = 0.25$$

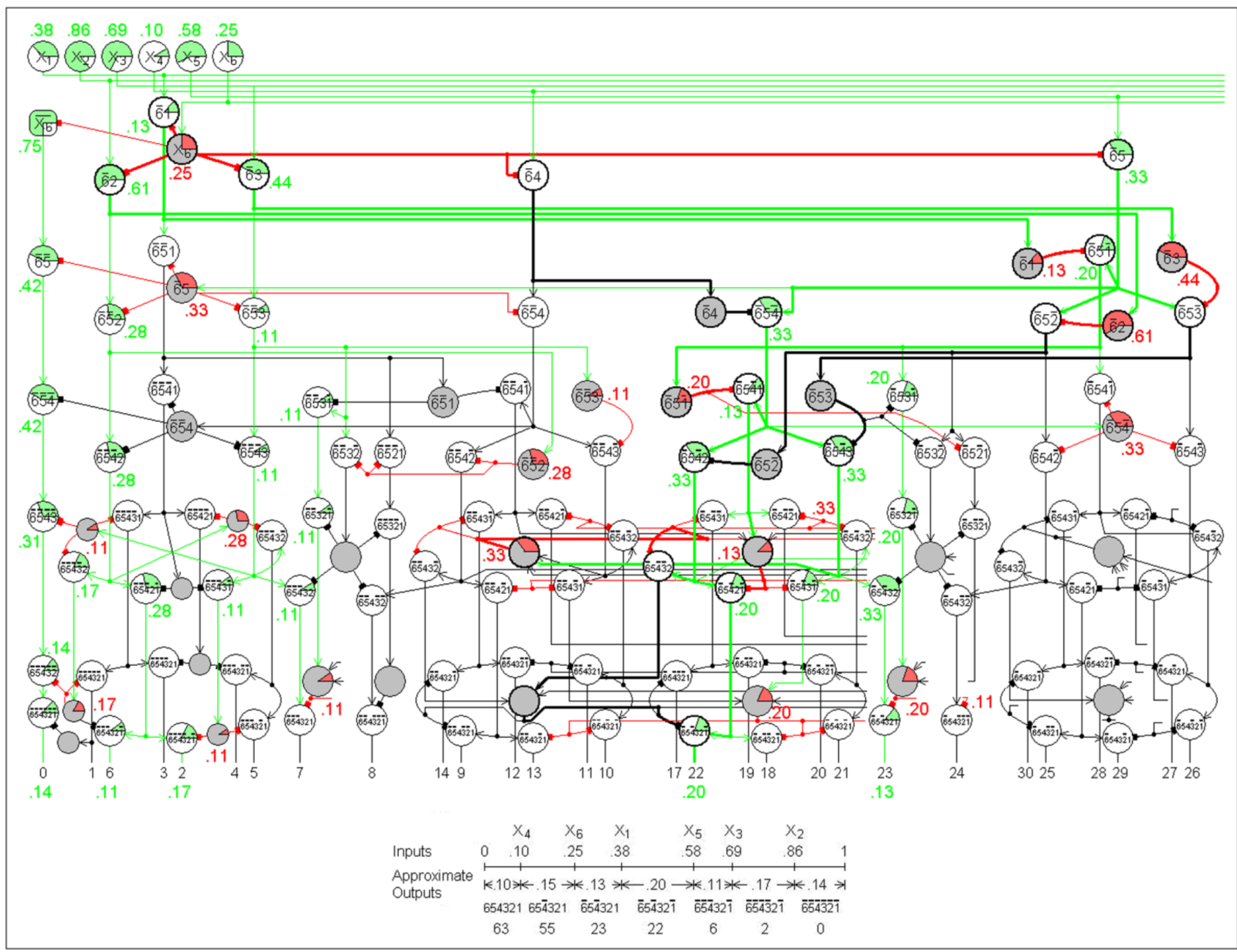

**Fig 11. A fuzzy decoder with shared inhibitory cells [5].** The circuit diagram shows nearly half of a fuzzy decoder with six inputs. The second and third columns can share inhibitory cells because the columns have mutually exclusive activity. To make the shared inhibitory connections clear, the figure omits outputs 15 and 16 that would otherwise be between these two columns. The example inputs show some cells in the inactive second column have inhibitory input from the third column, inhibiting them below the resting level. The line graph at the bottom shows the interval measure property of Table 1.

No information is lost in a fuzzy decoder; the original inputs can be recovered from the positive outputs and the interval measure property of Table 1. For example, taking the positive outputs in Fig 11 in order of the output number (right to left in the line graph), output 2 with X2 not negated implies X2 = 1 - output 0. Output 6 with X3 not negated implies X3 = X2 - output 2, output 22 with X5 not negated implies X5 = X3 - output 6, etc.

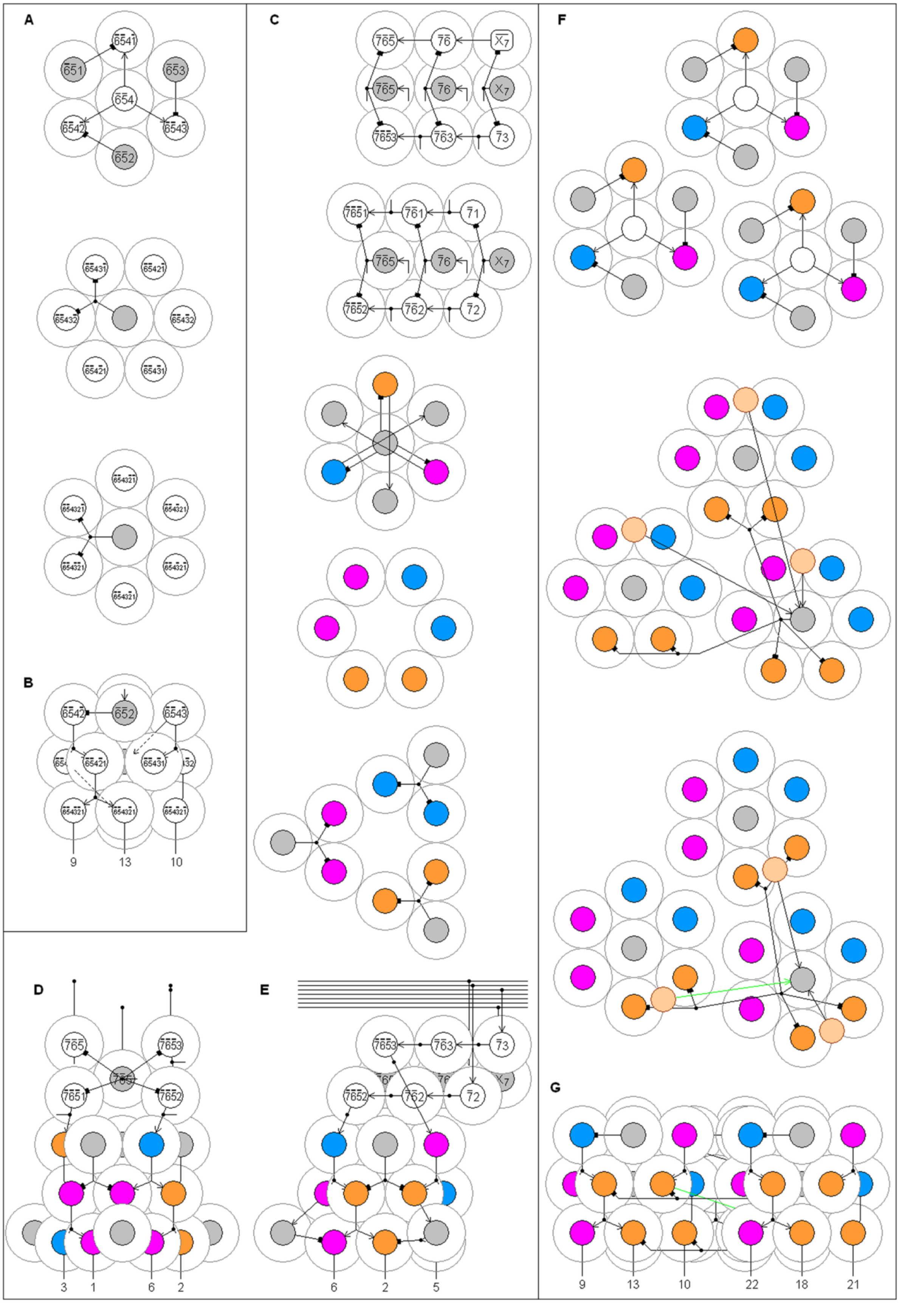

**Fig 12. Connection length and cell packing [5].** The connectivity of decoders suggests an arrangement of cells that optimizes total connection length and cellular packing density. Packing efficiency is illustrated with cell bodies contained in virtual spheres whose diameters are determined by the physiological constraints of necessary separations between cell bodies. Connection length is minimized when two connected cells are in adjacent spheres. The three layers of a typical circular column are shown in A, with the front view of the column in B. Five layers of cells that produce outputs 1–6 in a 7-input decoder are shown in C, with the side and front views in D and E. The lower three layers of three columns with shared cells are shown in F and G.

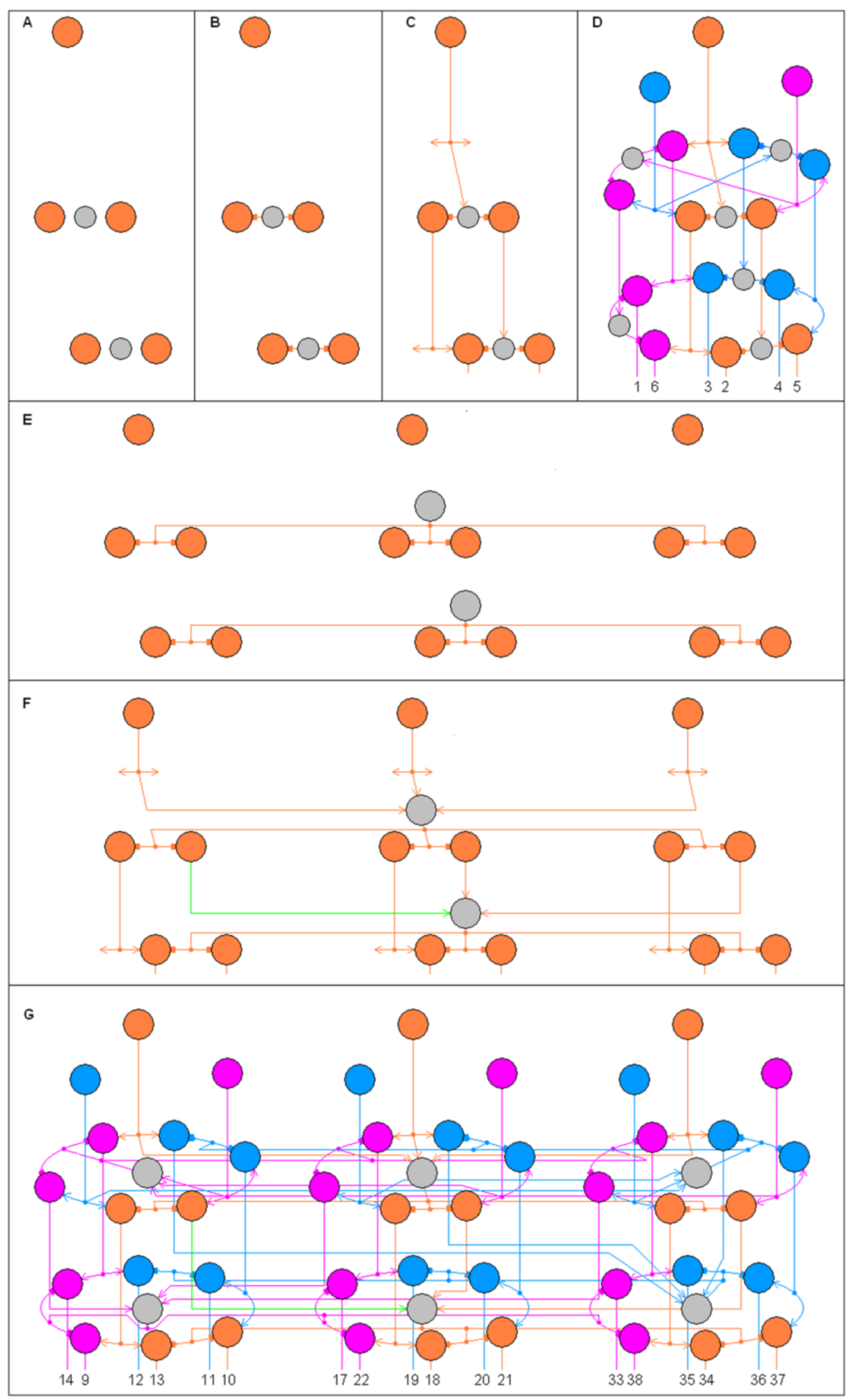

**Fig 13. Circular columns' modular architecture with repeated parts [5].** The logic identities in Table 1 that define the connectivity of the decoders make the decoders modular, recursive, and reductive. These features simplify construction during development or manufacturing. A column begins with unconnected cells in A. Inhibitory connections in B form identical three-cell parts in different layers. Excitatory connections in C combine these parts into a three-layer module. Three identical modules are joined in D to form a column. When several columns with mutually exclusive activity share inhibitory cells, as illustrated in Figs 11 and 12F and G, modular construction is nearly as simple. Inhibitory connections in E form identical parts in different layers, and excitatory connections in F combine these parts into a three-layer module. Additional excitatory connections in G join three of these modules to form three columns. The numbered outputs in G illustrate columns with mutually exclusive activity. The first two columns match the mutually exclusive columns in Fig 11.

The blue and violet versions of Fig 13F have the inhibitory cell on the left and right, respectively. Any number of columns in a set can share the same six inhibitory cells shown in Fig 13G simply by adding more modules to Fig 13F like the left and right sides of Fig 13F (and doing the same for the blue and violet versions of Fig 13F). The cell colors match in Figs 12 and 13.

It is not clear whether minimizing the number of cells as in Fig 9B minimizes total cost. The simplicity of the complete circular columns of Figs 8, 9, 11, 12, and 13 may reduce construction cost, as in Fig 13, and have other benefits such as reliability.

#### 4.4.2.1. Layers

The cerebral cortex is a thin sheet consisting of distinct layers. The most typical form consists of six layers. The outer layer is acellular, consisting mainly of dendrites from lower layer cells and axons that lie within this layer and make connections in other areas of the cortex. The white matter lies below the innermost layer [28, p. 345].

Fuzzy decoders exhibit all of these features. The layers are illustrated in Figs 7 and 8 and 10-13. Five cellular layers are sufficient for any number of inputs, and five are necessary if the number of inputs is at least seven [5]. The five cellular layers and acellular outer layer are shown most clearly in Fig 12E. The decoder inputs are to the outer two cellular layers, as shown in Figs 12C-E. So for minimized connection lengths, input axons should be close to these layers, i.e., in the top layer. The decoder outputs are in the lower layer, leading to the white matter.

Fuzzy decoders with few inputs can account for the structure of some other parts of the nervous system that have fewer layers, such as the three-layer retina, as demonstrated by the color vision model in Figs 6, 7, and 10B and D, and other parts of the brain that are phylogenetically older than the neocortex, such as the hippocampus and cerebellar cortex that both have three cellular layers, as demonstrated by the general three-layer decoder in Fig 8C.

**4.4.2.2. Columnar computational modules**

Information is processed in small columns of neurons connected across the cortical layers and spanning all layers. Each column is a fraction of a millimeter in diameter. Columns are thought to be the fundamental computation modules of the neocortex [28, p. 348].

Fuzzy decoders exhibit these properties. Separate circular columns are necessary for decoders with more than three inputs, as demonstrated in Fig 9C, and illustrated in Figs 8D, 11, 12F and G, and 13E-G. Each column is restricted to no more than six outputs, which means the columns must be small.

A fourth of a decoder's outputs are not produced by the circular columns, as illustrated in Figs 8C and D, 9C, and 11. These non-circular columns are limited to two outputs, so they are also small. The recursive and modular architecture of these non-circular columns is discussed in [5, Figure 3]. Only the circular columns are discussed here.

**4.4.2.3. Information transformation through the layers**

Information is transformed in each layer of the cerebral cortex while passing through the columns in series and in parallel [28, p. 570].

This property is illustrated by the example neuron output values in Figs 7, 8, 10, and 11.

#### 4.4.2.4. Suppressed regions

Active regions in the cerebral cortex are surrounded by neurons that are inhibited below the resting level [29].

Fuzzy decoders have sets of columns with mutually exclusive activity, i.e., when one column is active the others are inactive. Inhibitory cells can be shared by these columns with no appreciable effect on the inactive columns. Shared inhibitory cells have two consequences. They reduce the number of cells required, and they produce active regions that are surrounded by neurons inhibited below the resting level.

In Fig 11, every output cell in the second column contains NOT X5 AND X4. Every output cell in the third column contains X5 AND NOT X4. The interval measure property in Table 1 implies that one of these columns can have no positive output, depending on which of the two inputs X5 or X4 is larger. (If the two are equal, neither column has a positive output.) In a fuzzy decoder with five or more inputs, every column except two is a member of a set of mutually exclusive columns. As the number of inputs increases, the number of such sets and the number of columns in each set increases. For example, a decoder with 5 inputs has 4 columns with 1 set of 2 mutually exclusive columns; a decoder with 6 inputs (partially shown in Fig 11) has 8 columns with 2 sets of 3 mutually exclusive columns; a decoder with 7 inputs has 16 columns with 2 sets of 4 mutually exclusive columns and 1 set of 6 mutually exclusive columns.

Mutually exclusive columns can share inhibitory cells without having an appreciable effect on the inactive column. Shared inhibitory cells are illustrated in Figs 11, 12F and G, and 13E-G. Each shared inhibitory cell inhibits two cells in each inactive column in the same set (Fig 13F). If connection length is minimized, most columns are surrounded by several columns in the same set.

Inhibition below the resting level occurs if a cell receives inhibitory input with no excitatory input. When one column in a set of mutually exclusive columns is active, the shared inhibitory cells suppress cells below the resting level in the surrounding inactive columns.

**4.4.2.5. Placement and proportion of inhibitory cells**

Inhibitory cells are in every cellular layer and constitute 20 to 25% of the neurons in the neocortex [28, p. 348].

Figs 8 and 11-13 show that fuzzy decoders have inhibitory cells in all cellular layers.

The second consequence of shared inhibitory cells, mentioned in the previous section, is fewer inhibitory cells. Fig 14 shows the proportion of inhibitory cells in fuzzy decoders is 20.4% to 25% if the number of inputs is at least six [5]. This closely matches the 20 to 25% inhibitory cells in the cerebral cortex. Without sharing of inhibitory cells, the percentage of inhibitory cells in fuzzy decoders is considerably more than 25%. So fuzzy decoders that minimize resource requirements by sharing inhibitory cells agree well with empirical evidence in the brain.

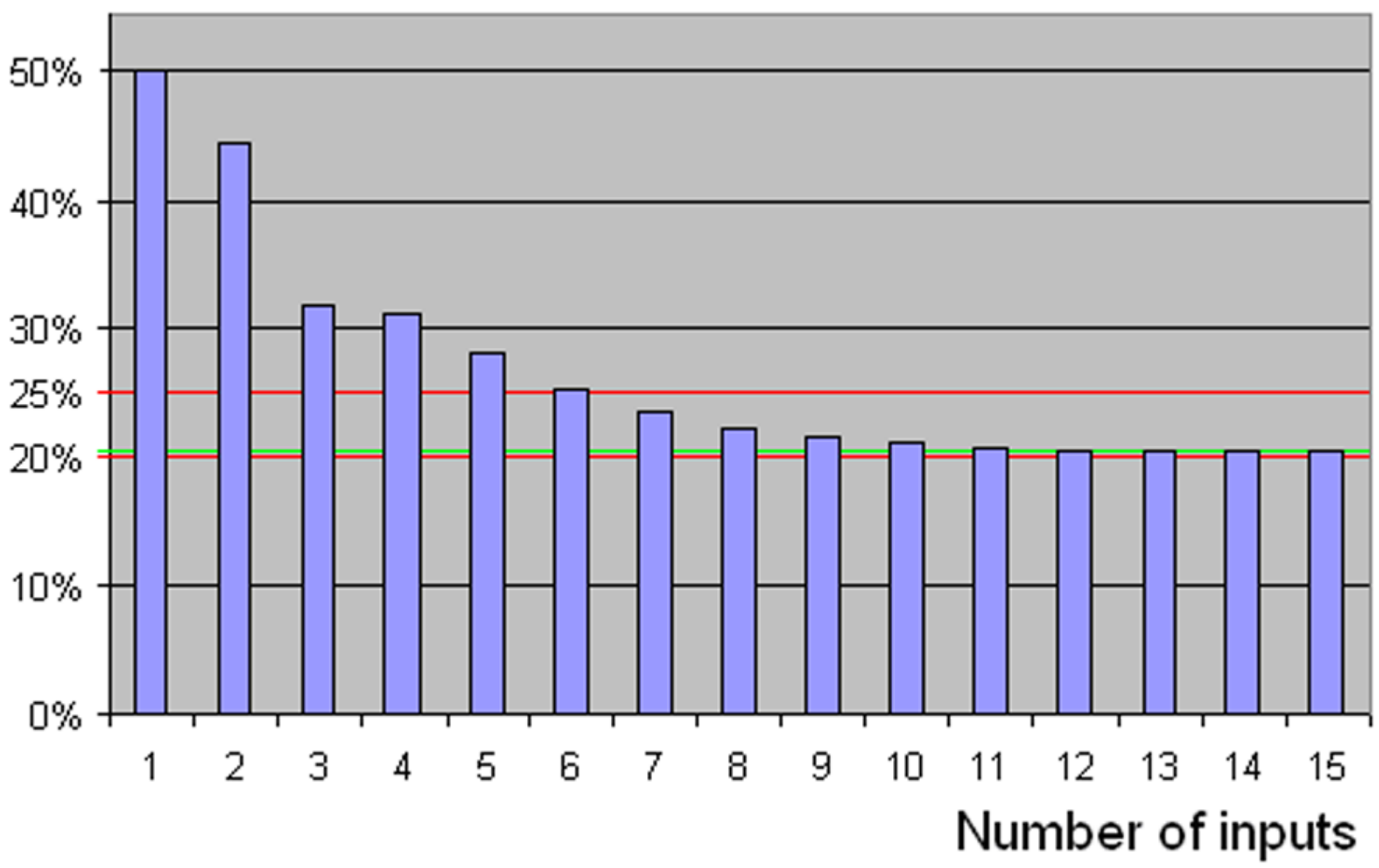

**Fig 14. Portion of fuzzy decoder cells that are inhibitory [5].** The proportion of inhibitory cells in fuzzy decoders is 20.4% to 25% if the number of inputs is at least six

#### 4.4.2.6. Connection directions

Weigert stains for myelinated fibers show that nearly all axons run parallel or perpendicular to the layers [30]. Only a few axons run diagonally for short distances, no more than from one layer to the next. Weigert stains also show nearly all axons in the outer two layers (including the acellular layer) and much of the third layer are parallel to the layers.

These restrictions on connection directions are not an insignificant feature of the brain, considering the possible directions in three-dimensional space. Figs 8, 11, and 12 show these horizontal and vertical connections in fuzzy decoders. For six or more inputs, Figs 11 and 12 show that fuzzy decoders match the above general properties found in the Weigert stains almost precisely. They show that nearly all connections in the lower three layers of a fuzzy decoder are horizontal or vertical. They also show that nearly all axons in the outer two layers (including the acellular layer) and much of the third layer are parallel to the layers (Fig 12E). The only diagonal connections are the excitatory inputs from one column to shared inhibitory cells in another column (Fig 12F). These connections extend only from one layer to the layer below (Fig 13E-G).

### 4.5. Short-term memory derived from a standard logic circuit by the AND-NOT transform

The networks proposed in [7] show how neurons can be connected to form flip-flops, which are standard memory mechanisms in electronic computational systems and the basic building blocks in sequential logic systems. A neuronal flip-flop (NFF) can be derived from certain electronic designs by a simple transformation. This process is illustrated in Fig 15. Fig 15D shows a standard electronic flip-flop design. Fig 15E shows a flip-flop that can be implemented with two neurons. Either design can be derived from the other by the transformation of moving each negation circle from one end of the connection to the other. If a circle is moved past a branch point to an output, the output is inverted. The transformation does not affect the logic of the network, but it changes the logic of each component in Fig 15D to a logic function that can be implemented by a single neuron in Fig 15E. Although the transformation is a minor modification, both the transformation itself and the resulting networks that contain AND-NOT gates are likely to be new to engineering because the AND-NOT gate is virtually never used as a building block in electronic logic circuit design.

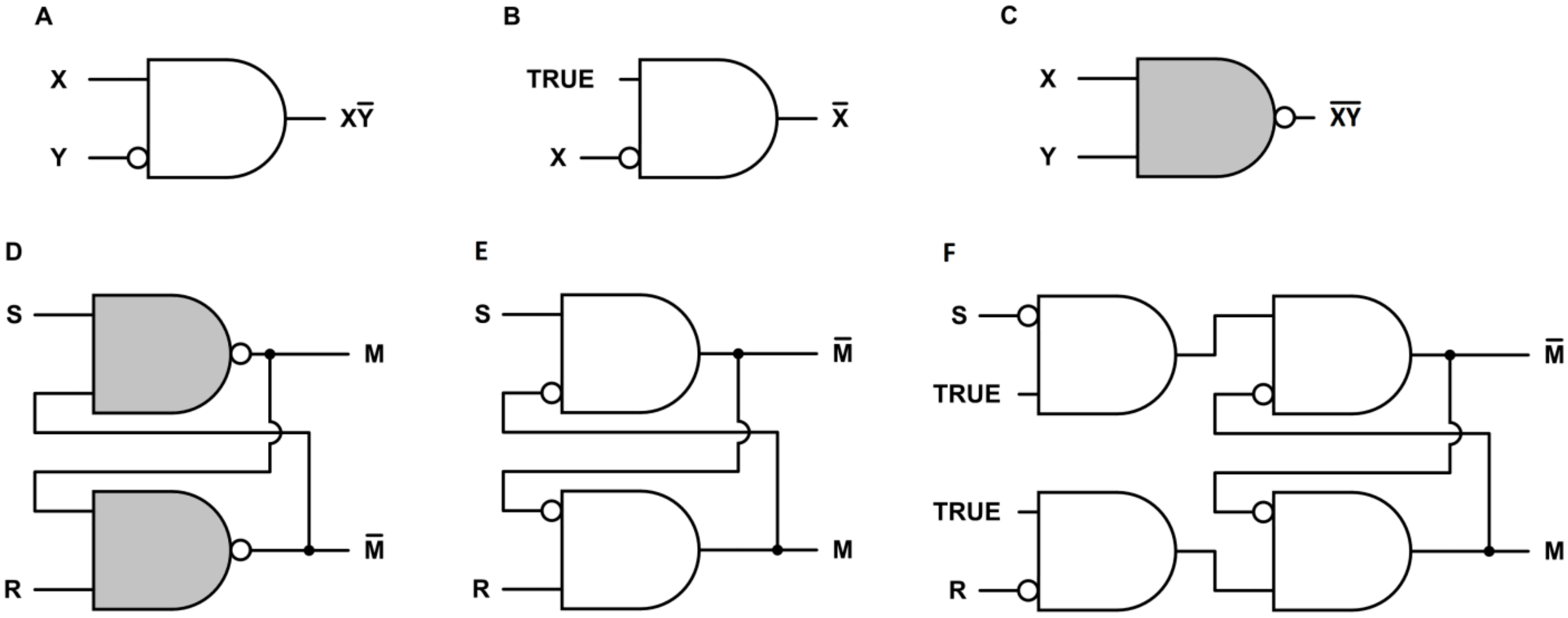

**Fig 15. Neural logic gates and flip-flops [7]. A.** A symbol for an AND-NOT logic gate from Fig 1, with output X AND NOT Y. The symbol can also represent a neuron with one excitatory input X and one inhibitory input Y. **B.** An AND-NOT gate configured as a NOT gate, or inverter. **C.** A NAND gate (NOT AND). The output is NOT (X AND Y). There is no obvious way to implement this gate with a single neuron. **D.** A standard design for an electronic active low Set-Reset (SR) flip-flop composed of two NAND gates. **E.** An active low Set-Reset (SR) flip-flop composed of two AND-NOT gates. This design is derived from the design in D by moving each negation circle from one end of the connection to the other. **F.** An active high SR flip-flop.

Figs 15D and E show active low set-reset (SR) flip-flops. The S and R inputs are normally high. A brief low input S sets the memory state to M = 1, and a brief low input R resets it to 0. Feedback maintains a stable state. Inverting the inputs of Fig 15E produces the active high SR flip-flop of Fig 15F. The S and R inputs are normally low. A brief high input S sets the memory M to 1, and a brief high input R resets it to 0.

Fig 16 shows a simulation of the NFF in Fig 15F, using the noise-reducing neuron response function of Fig 4.

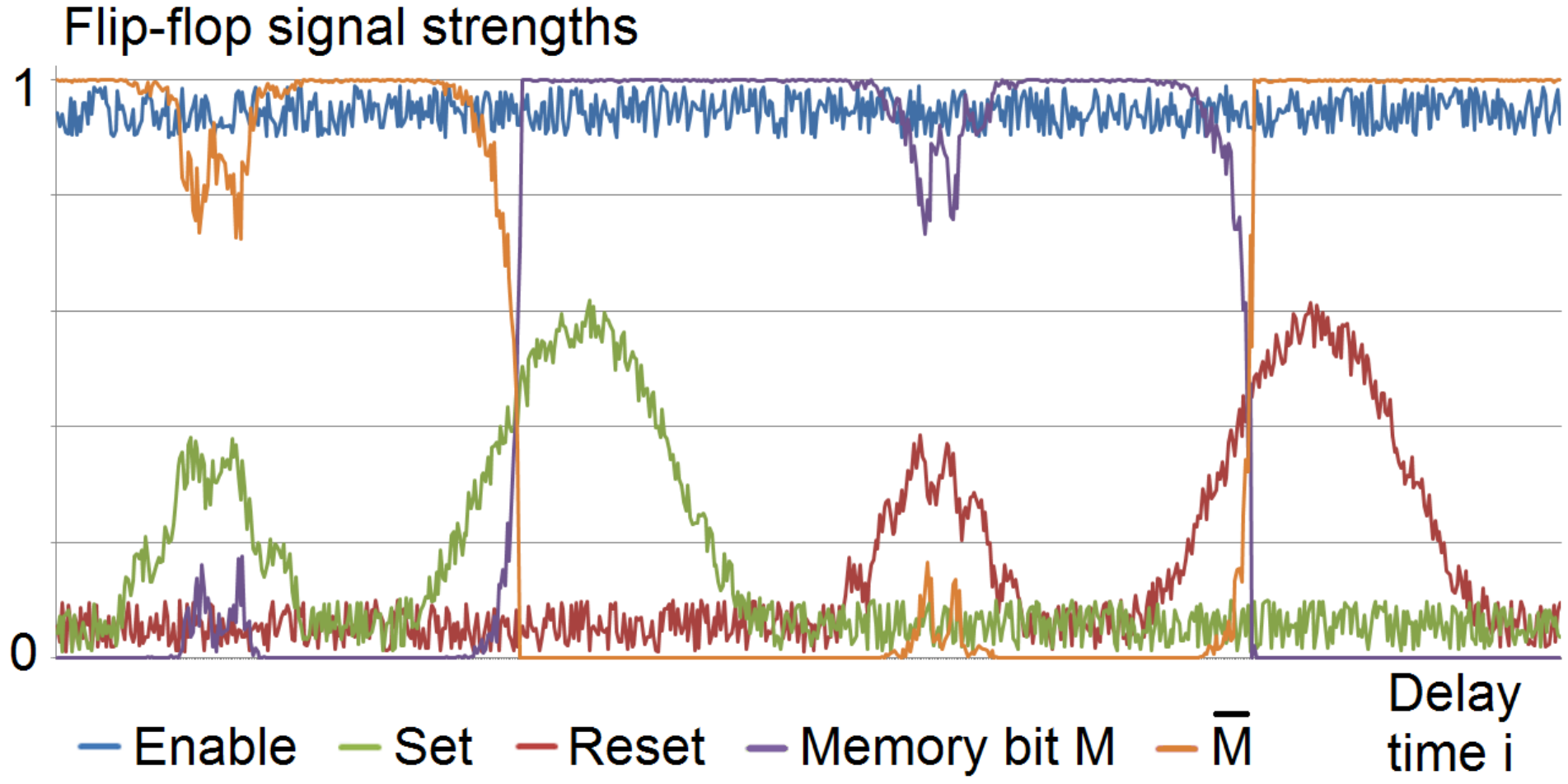

**Fig 16. Simulation of an NFF operation with noise in the inputs [7].** This simulation of the NFF in Fig 15F shows the NFF's operation is robust in the presence of moderate levels of additive noise in binary inputs. The effect of baseline noise on the memory bit is negligible, and temporary bursts of larger noise have no lasting effect.

Memory tests have shown that certain neurons fire continuously at a high frequency while information is held in short-term memory. These neurons exhibit seven characteristics associated with memory formation, retention, retrieval, termination, and errors [31-33]: 1) Before the stimulus was presented, the sampled neuron discharged at a low, baseline level. 2) When the stimulus was presented, or shortly after, the neuron began to fire at a high frequency. 3) The high frequency firing continued after the stimulus was removed. 4) The response was still high when the memory was demonstrated to be correct. 5) The response returned to the background level shortly after the test. 6) In the trials where the subject failed the memory test, the high-level firing had stopped or 7) had never begun.

It was shown in [7] that the NFF in Fig 15F produces all of these characteristics. The NFF also predicts eight more characteristics that can be tested by the same methods used in finding the first seven [7].

### 4.6. Brain waves from synchronization

Synchronization of neural structures' state changes to avoid timing errors is an advantageous function for information processing in the brain as well as in electronic computational systems. An oscillator can synchronize state changes by enabling the changes simultaneously. EEGs show widespread rhythms, commonly known as brain waves, that consist of many neurons firing with matched periods. Synchronization can produce this activity. Synchronization as the basis for brain waves was apparently first proposed in [8].

The trade-off for greater processing speed is a higher error rate. The relative importance of speed and accuracy and the complexity of information processing in a particular brain function determine the optimal frequency of state changes. Being a large processor with many functions, the brain would require synchronization for a wide variety of processing speeds, with some information requiring processing as fast as possible. Cascaded oscillators can provide synchronization with these two properties. Cascaded oscillators accurately produce the entire multimodal distribution of brain wave frequencies as a function of the mean and variance of neuron delay times [8].

#### 4.6.1. JK flip-flop or toggle

A problem with the SR flip-flops in Fig 15 is that an error occurs if both S and R activate simultaneously. The so-called JK flip-flop in Fig 17 solves this problem. The JK flip-flop is the most widely used flip-flop design in modern electronic computational systems because of its several advantageous features.

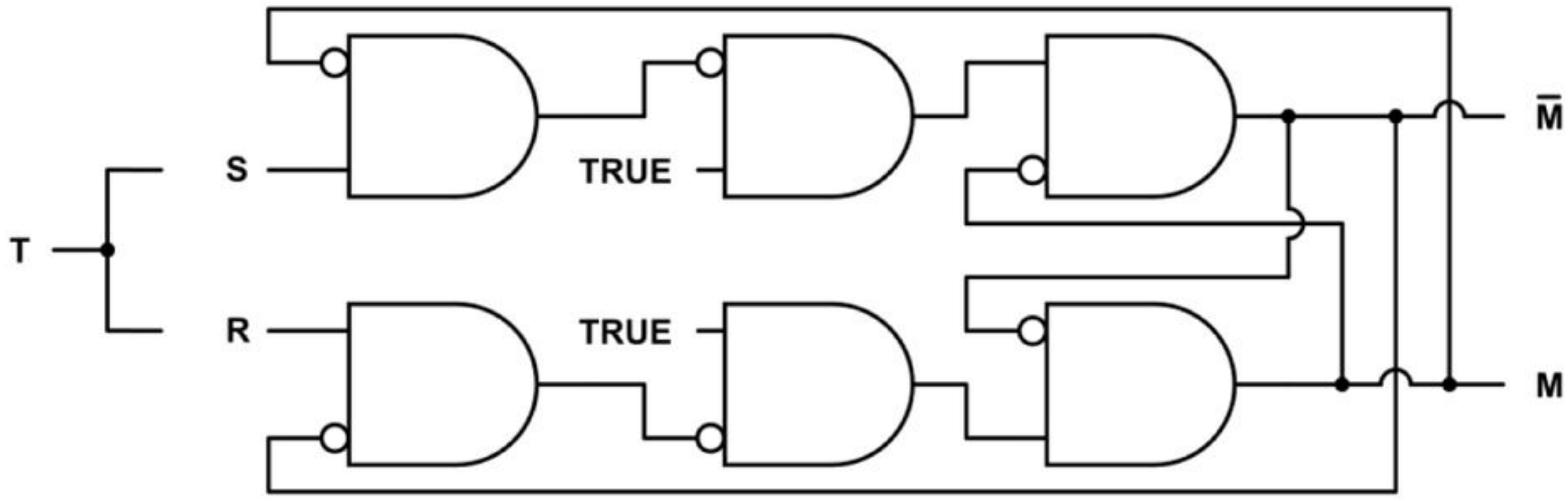

**Fig 17. JK flip-flop or toggle [8].** The network composed of AND-NOT gates can be implemented with neurons or electronic components. Here the S and R inputs to Fig 15F have been inhibited by the flip-flop outputs. If S and R are both high simultaneously, the flip-flop state is inverted because the most recent inverting input is inhibited by one of the outputs. This means the JK flip-flop can be configured as a toggle by linking the Set and Reset inputs, as illustrated by the single input T in the figure. Each time the input T is high, the toggle state is inverted.

**4.6.2. Cascaded oscillators**

An oscillator connected in sequence with toggles makes a cascade of oscillators. A toggle with input from an oscillator will oscillate at exactly double the period of the oscillator. This is because the oscillator must have a high output once to make one toggle output high and another high output to make the toggle output low.

An odd number of three or more inverters connected sequentially in a ring produces periodic bursts as each gate inverts the next one. Fig 18 shows a three-inverter ring oscillator connected in sequence with two JK toggles, forming a cascade of three oscillators.

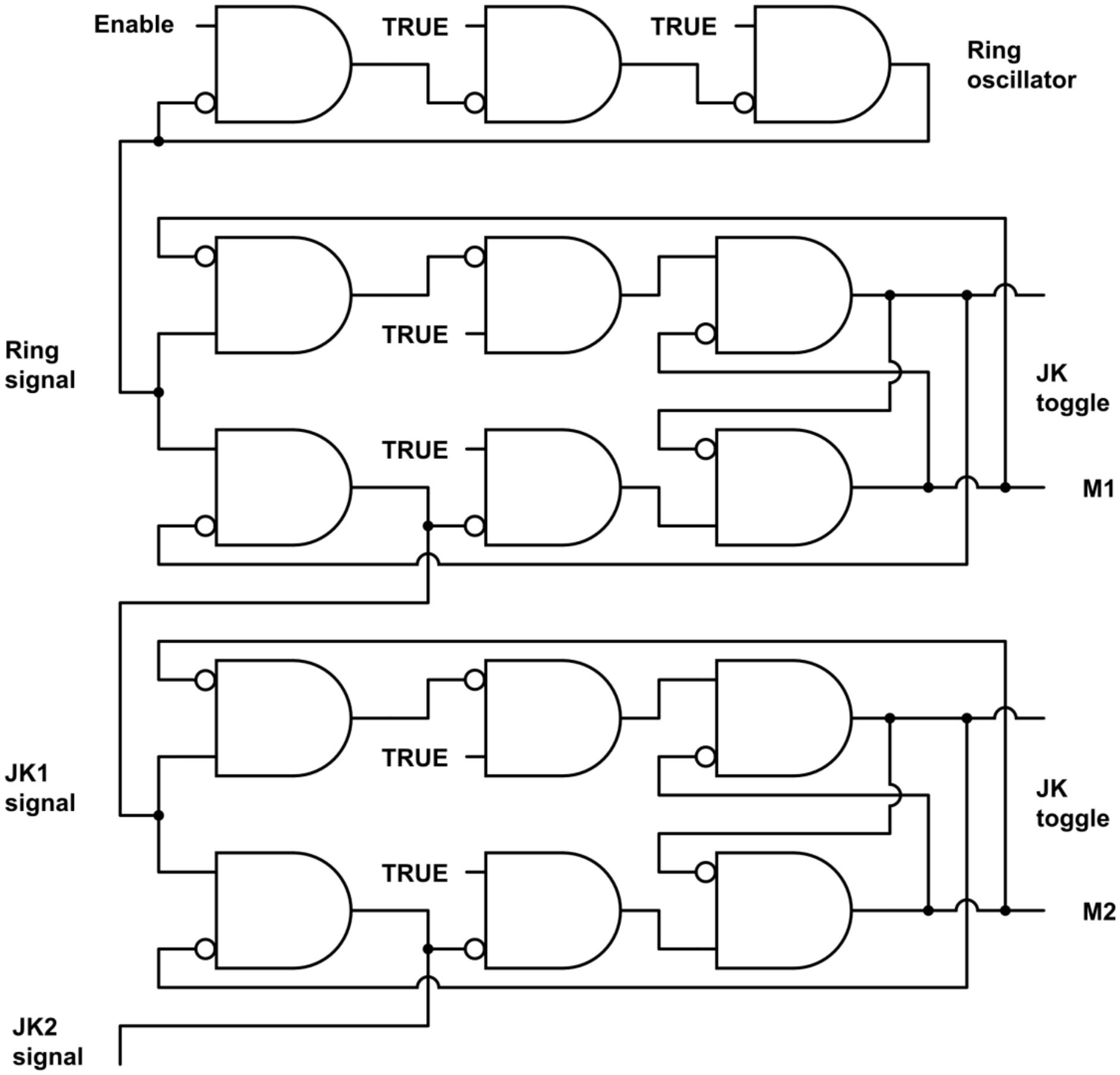

**Fig 18. Three cascaded neural oscillators [8].** The cascade consists of a ring oscillator followed by two JK toggles connected in sequence. The ring oscillator is composed of three inverters of Fig 15B, with an enabling first input. The toggles are as in Fig 17. One of the first cells in each toggle is used for the output signal because it maintains the burst duration of the input.

Cascaded oscillators satisfy the need for some information to be processed as fast as possible because the three-inverter ring oscillator is the fastest oscillator that can be constructed with logic gates. The cascade also provides a wide variety of speeds because the oscillation period grows exponentially (doubling) with each successive oscillator in the cascade. The brain would not need to be limited to one cascade of oscillators. Several cascades could handle different computational functions.

### 4.6.3. Distribution of brain wave frequencies

EEGs show widespread rhythms, commonly known as brain waves, that consist of many neurons firing with matched periods. Brain waves have a somewhat complex distribution of frequencies. The frequencies are widely distributed across five bands. Each band has a single mode that is not near the midpoint of the band's endpoints. The modes and endpoints have octave relationships, with each being double the frequency of the mode or endpoint of the preceding band. None of these properties has an apparent function. Cascaded oscillators produce all of these phenomena as by-products of the design for the useful properties of synchronization.

Cascaded oscillators as shown in Fig 18 accurately produce all of the specific mode and endpoint frequencies of brain waves as explicit functions of just two parameters, the mean and standard deviation ($\mu_d$ and $\sigma_d$) of neuron delay times in the ring oscillators. For example, the boundary separating the alpha and beta frequency bands is

$$125/\{\mu_d + \sqrt{[(\mu_d)^2 + (\sigma_d)^2 \ln(4)]}\} \text{ hertz.}$$

With 4 and 1.5 ms being the best estimates for $\mu_d$ and $\sigma_d$, respectively, this predicted boundary value is 14.9 Hz. This and the other predicted values are shown in Fig 19.

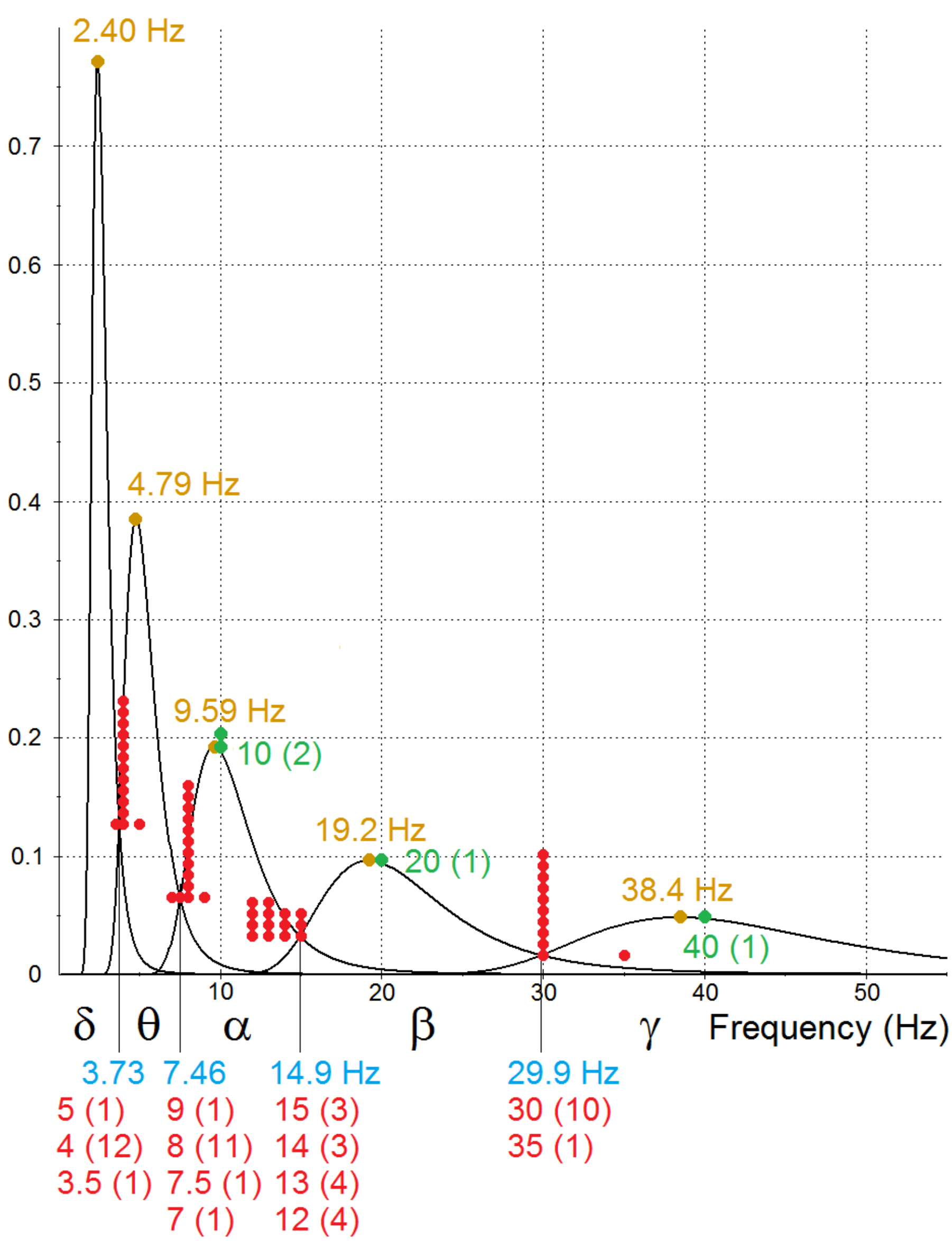

**Fig 19. Distribution of EEG frequencies compared to predictions by cascaded oscillators [8].** This figure shows how closely the activity of neural structures synchronized by cascaded oscillators fits the oscillations found in EEGs. Frequencies that are commonly cited as partition points separating the EEG delta, theta, alpha, beta, and gamma frequency bands and peak frequencies of three of the bands are shown in red and green, respectively. Numbers in parentheses and numbers of data points show how many times each frequency was found cited. The graphs are the estimated probability distribution functions (PDFs) of the frequencies of a three-neuron ring oscillator and four cascaded toggles. All of the PDFs were determined by the estimated mean and variance of neuron delay times. The five intervals defined by the intersections of consecutive PDFs are labeled with Greek letters to distinguish them from EEG frequency bands, which are often written in the Roman alphabet. The intersections and modes are labeled in blue and yellow, respectively.

Ring oscillators, JK flip-flops, and cascaded toggles as oscillators are not new to engineering. What is new is the design with components that can be implemented with neurons and the resulting frequency distribution that closely matches the distribution of brain wave frequencies. This distribution in five bands is a function of only the mean and variance of neuron delay times in the first oscillators in the cascades.

#### 4.6.4. JK toggle weaknesses and epileptic seizures

The cascade design's necessarily narrow tolerance for error suggests a possible relationship to the abnormal electrical activity characteristic of epileptic seizures. The JK toggle was chosen for the cascade design because it can function correctly with input from a three-inverter ring oscillator and the long bursts from toggles later in the cascade. Other toggles are either too slow to function correctly with input from a three-inverter ring oscillator, or too fast for input of long bursts from later toggles.

The JK toggle is not without weaknesses. To function correctly as a toggle, it has narrow tolerances for errors in the input frequency and the input burst duration. The resulting synchronization errors in neural firing and the brain's efforts to deal with them may be related to the abnormal electrical activity in neurological disorders such as epileptic seizures [8].

Configuring JK toggles as cascaded oscillators may be new to engineering because of the limitations of neurons. For example, although the three-inverter ring oscillator may be the fastest oscillator that can be constructed with logic gates, electronic devices have access to oscillators that are not confined to logic gates.

#### 4.6.5. Longstanding controversy of short-term memory

Experiments have found short-term memory to be associated with some neurons firing persistently and others firing in brief, coordinated bursts. Which phenomenon is predominant in memory has been a longstanding controversy [34-35]. Together, the NFFs in Fig 15 and synchronization by cascaded oscillators in Fig 18 suggest a resolution to the controversy: Memory is stored by persistent firing in flip-flops, and the coordinated bursts observed along with the persistent firing are due to the stored information being processed by several neural structures whose state changes are synchronized by a neural oscillator [8]. An example of such short-term memory processing is a telephone number being reviewed in a phonological loop.

### 4.7. Stomatogastric ganglion

The stomatogastric ganglion (STG) is a group of about 30 neurons that resides on the stomach in decapod crustaceans. Its two central pattern generators (CPGs) control the chewing action of the gastric mill and the peristaltic movement of food through the pylorus to the gut. The STG has been studied extensively because it has properties that are common to all nervous systems and because of the small number of neurons and other features that make it convenient to study. The organization of synaptic connections has been completely mapped [9, Fig 4]. So many details are known that the STG is considered a classic test case in neuroscience for the reductionist strategy of explaining the emergence of macro-level phenomena from micro-level data [2, p. 4]. Despite the intense scrutiny the STG has received, how it generates its rhythmic patterns of bursts remains unknown [2, p. 5].

The oscillator designs in [9] show how the STG's rhythms can be produced. The oscillators also share several major features of the STG. One of the designs, discussed here, fills a major gap in the well-known family of inverter ring oscillators and is apparently new to engineering.

### 4.7.1. The American lobster's four rhythms

The movements of the muscles that produce the peristaltic movement of food through the pylorus to the gut is usually described as a triphasic pattern. The pyloric rhythms in the American lobster (*Homarus americanus*), however, were shown to have four distinct oscillations, with phases approximately uniformly distributed over the common period [36].

### 4.7.2. Inverter ring oscillators

An inverter, or NOT logic gate, is shown in Fig 15B. The output is the opposite of the single variable input. As discussed above, an inverter ring oscillator consists of an odd number of inverters connected sequentially in a ring. It is a standard design in electronic computational systems. An example composed of three inverters is shown in Fig 18. The inverter ring oscillator produces an odd number of oscillations, with identical frequencies, and phases uniformly distributed over the period. An inverter ring oscillator therefore could not produce the four phases of the American lobster.

### 4.7.3. Flip-flop ring oscillators

A flip-flop ring oscillator is composed of two or more flip-flops connected in a ring, with both outputs of each flip-flop connected as inputs to the next one. Fig 20 shows four flip-flop ring oscillators.

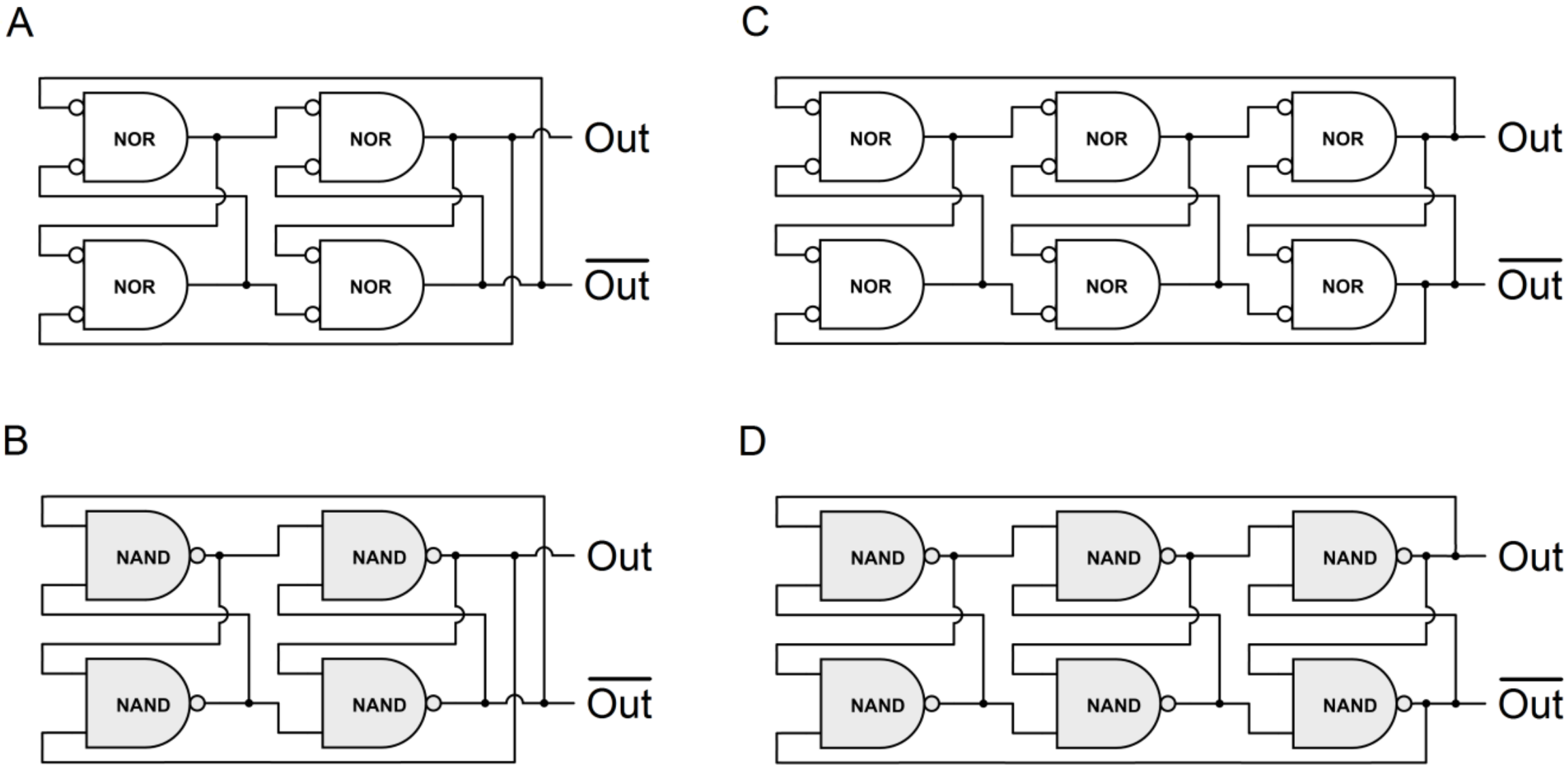

**Fig 20. Flip-flop ring oscillators [9].** Each network is composed of two or three simple flip-flops, connected sequentially to form a ring. The rings in Figs B and D are composed of flip-flops in Fig 15D. Each white network performs the same function as the gray network below it. Each network in each pair can be derived from the other by the transformation of moving the negation circles from one end of a connection to the other. All four designs are apparently new to both neuroscience and engineering.

Both NOR (NOT OR) and NAND gates are commonly used in electronic logic circuits, but the NOR gate can be implemented by a neuron with two inhibitory inputs if the neuron is spontaneously active or has continuous excitatory input. This follows from the same argument used for the neuron's AND-NOT capability: Since one inhibitory input can suppress an excitatory input (or a spontaneously active neuron), either one of two inhibitory inputs can suppress it. That is, the output is high when neither inhibitory input is high. Such gates are shown in Figs 20A and 20C. The gate's Boolean output is NOT X AND NOT Y. This means the neuron can function as a NOR gate. That is, NOT (X OR Y) = NOT X AND NOT Y by De Morgan's law.

The flip-flop ring oscillator is an extension of the inverter ring oscillator. It is apparently new to engineering, and it fills a major gap in the family of ring oscillators: An inverter ring oscillator produces an odd number of three or more oscillations, and a flip-flop ring oscillator produces an even number of four or more oscillations. Both types of ring oscillators produce oscillations with identical frequencies, with phases uniformly distributed over the period.

The flip-flop ring must be connected differently depending on whether the number of flip-flops is even or odd. As shown in Fig 20, the last component in each row is connected to the first component in the other row if the number of flip-flops is even, and connected to the same row if the number is odd. That is, the even number of flip-flops is connected like a Möbius strip, and the odd number like a ring.

### 4.7.4. Comparison of lobster and flip-flop ring oscillations

Fig 21 shows a comparison of the American lobster's pyloric rhythms and simulated oscillations of a flip-flop ring composed of neurons. The simulation image size was adjusted to match the period with the image of the lobster's oscillations, and the centers of the first bursts were aligned (green and blue).

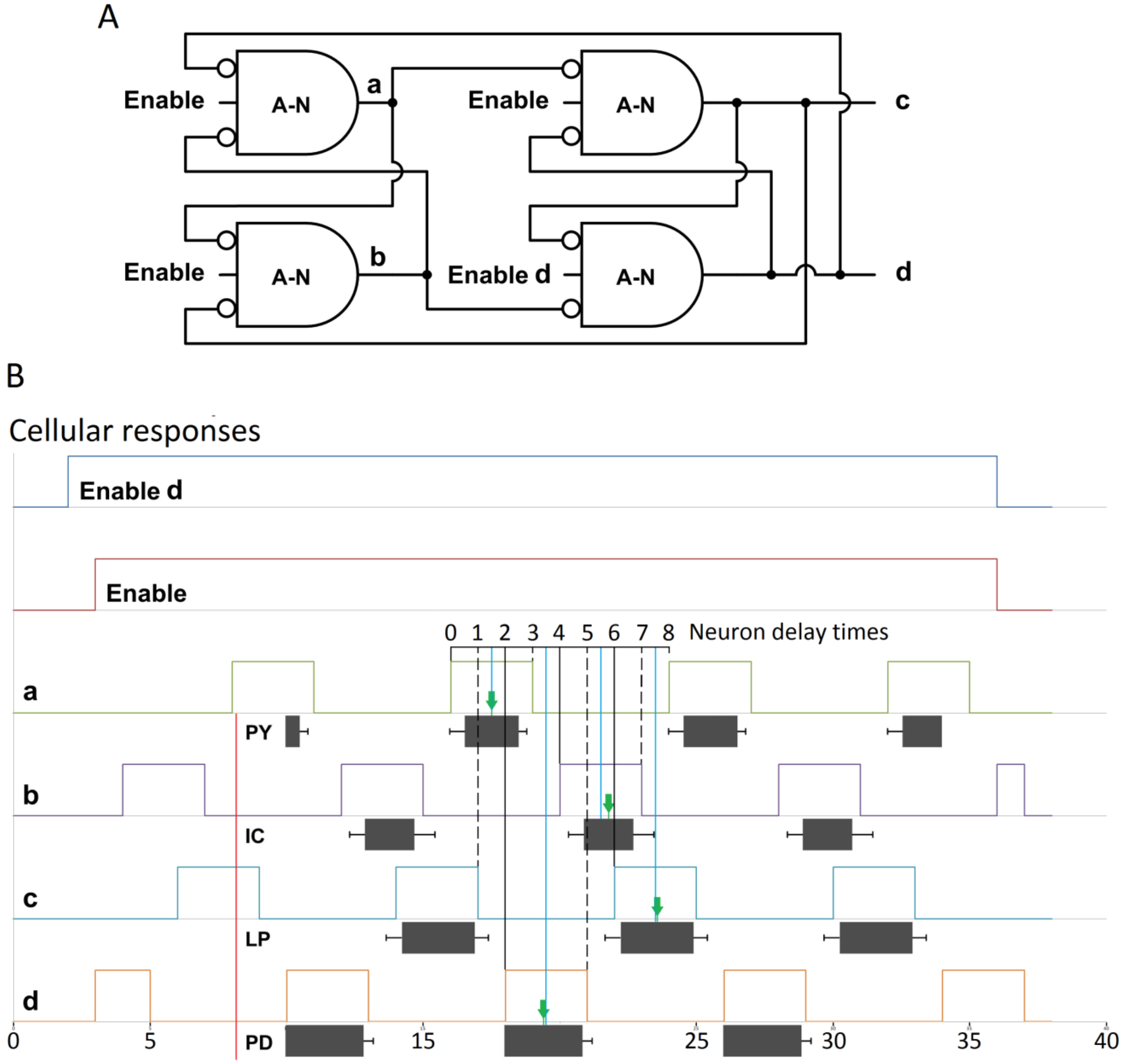

**Fig 21. Comparison of lobster and flip-flop ring oscillations [9] A.** The flip-flop ring oscillator of Fig 20A with excitatory enabling inputs from the brain or sensory cells. **B.** A comparison of a simulation of the flip-flop ring oscillator, shown in square waves, to the averages of 17 recordings of the American lobster's four pyloric oscillations, shown in blocks with standard deviation bars (adapted from [36]). The period of each of the flip-flop ring's oscillations is eight neuron delay times. Their phases are uniformly distributed over the period at both the burst onsets, indicated by black lines, and at the burst centers, indicated by blue lines.

The black lines and dashed lines together show that at each delay time, one of the four gates changes states. The green arrows indicate the centers of the pyloric bursts.

The close fit between the blue lines and green arrows shows that the oscillator and pyloric phases, as determined by the burst centers, are approximately the same. The burst durations are also close, especially when the standard deviation bars are considered. In principle, a CPG can generate any number of oscillations, with any frequencies, phases, and duty cycles (ratio of high signal duration to period). So the overall fit between flip-flop ring and lobster CPG is quite close.

The flip-flop ring in Fig 21A is composed of NOR gates with excitatory inputs, instead of spontaneously active NOR gates as in Fig 20A, so the cells can be enabled by excitatory input as needed. When enabled, the neurons function as NOR gates. The enabling input could be from the brain or sensory receptor cells. Nearly all STG cells have at least one excitatory input (Fig 4 in [9]).

#### 4.7.5. Similarities between the models and the STG

The neural networks proposed in [9], including the flip-flop ring oscillator in Fig 21A above, model the pyloric CPG of the American lobster (*Homarus americanus*). The models share enough significant features with the lobster's pyloric CPG (Fig 4 in [9]) that they may be considered first approximations, or perhaps simplified versions, of STG architecture. The similarities include 1) mostly inhibitory synapses; 2) pairs of cells with: reciprocal, inhibitory inputs; complementary outputs that are180 degrees out of phase; and state changes occurring with the high output changing first; 3) coordinated oscillations with the same period and four phases distributed approximately uniformly over the period; 4) a close fit in duty cycles; and 5) at least one excitatory input to nearly all cells. All of these similarities are by-products of the flip-flop ring oscillator that was designed only to produce four oscillations with different phases.

## 4.8. Lamprey locomotion

The lamprey is one of the most ancient of extant vertebrate species. The lamprey's primitive nervous system has been studied extensively, and the basic architecture of the portion that produces its undulatory swimming motion is well known [37-42]. How the network produces this motion, however, remained unknown until recently. It was shown in [10] that each segmental component of the lamprey's CPG is a JK flip-flop, with additional excitatory inputs and feedback that cause all of the neurons' states to oscillate. As stated above, the JK flip-flop is the most widely used flip-flop design in modern electronic computational systems. This is apparently the first discovery of how a network of neurons with known connectivity functions as a logic circuit. Moreover, the lamprey's oscillator design is apparently new to engineering. The lamprey's use of the JK flip-flop design is a remarkable convergence of evolutionary design and modern technological design more than 400 million years apart.

### 4.8.1. JK flip-flop

Fig 22 shows two implementations of the JK flip-flop. The flip-flop in Fig 22A is logically the same as Fig 17. The two NOT gates in Fig 17 are replaced with negation circles in Fig 22A. (The extra cells in Fig 17 were necessary for timing.)

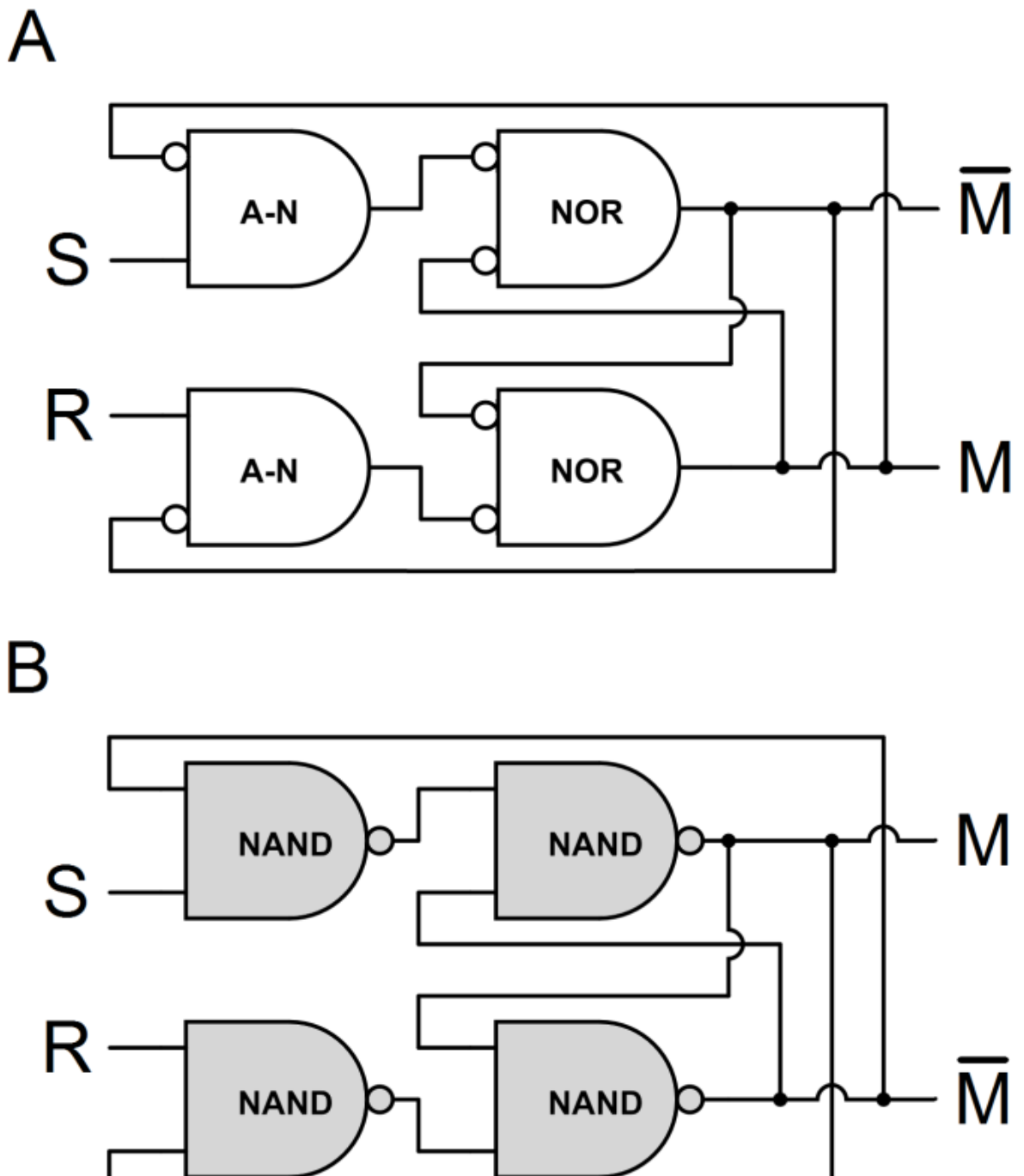

**Fig 22. JK flip-flops [10]. A.** A JK flip-flop composed of logic gates that can be implemented with neurons or with electronic components. **B.** One of the standard designs for an electronic JK flip-flop. Each design can be derived from the other by the transformation of moving each negation circle from one end of a connection to the other as discussed above.

### 4.8.2. Lamprey segmental controller of swimming motion

Fig 23A shows a schematic of the segmental component that controls the lamprey's undulatory swimming motion. Fig 23B illustrates the component in standard engineering form. This figure is a modification of the JK flip-flop in Fig 22A.

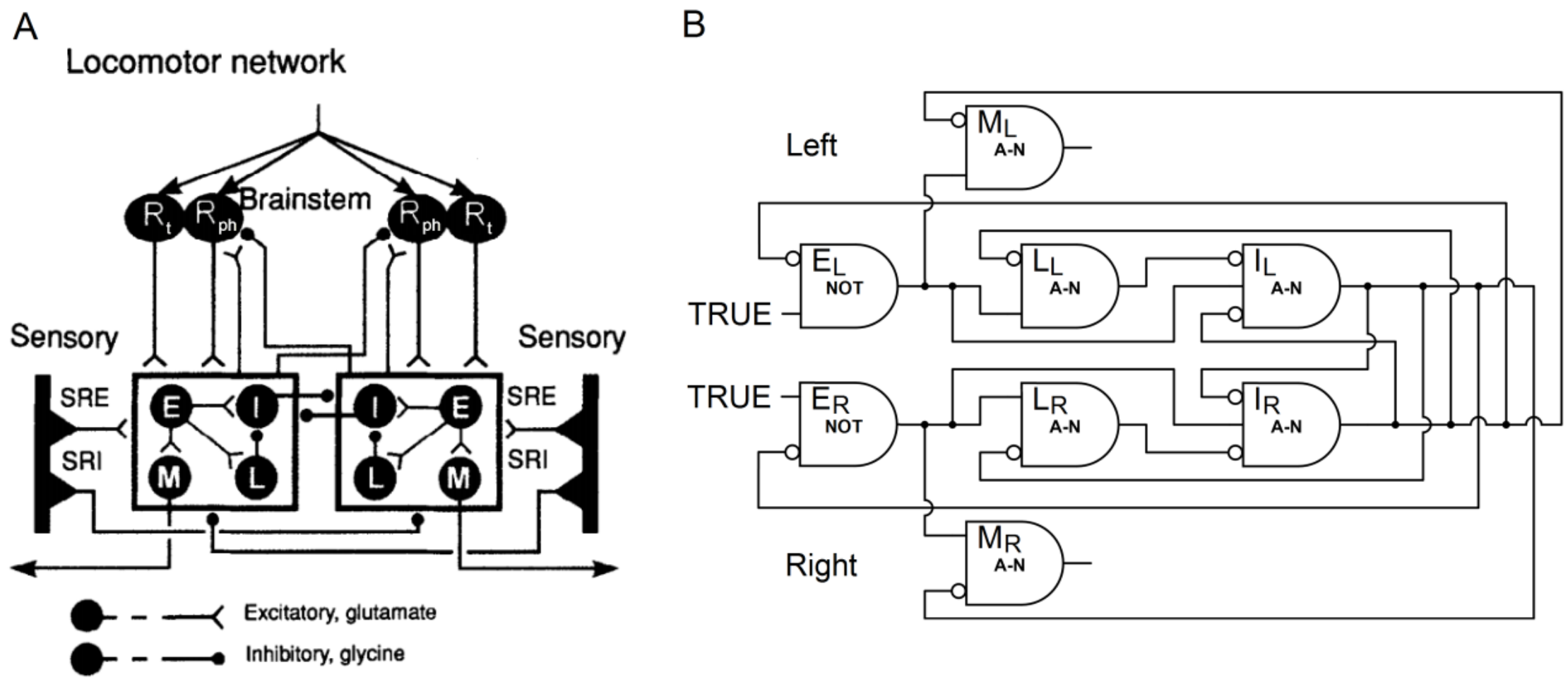

**Fig 23. Lamprey CPG that coordinates locomotion [10]. A.** The brainstem, sensory, and segmental components that generate bursts (adapted from [41]). A single segmental component is shown in the left and right squares. Each neuron symbol represents a group of neurons. Synapses that terminate at a square affect all neurons in the square. The excitatory neuron (E) in each square excites all of the other neurons in that square. The motor neuron (M) controls the muscles on one side of the segment. The commissural inhibitory neuron (I) inhibits all of the neurons in the contralateral square. The lateral neuron (L) inhibits the ipsilateral commissural neuron. The reticulospinal brainstem neurons are phasic (Rph) and tonic (Rt). The sensory neurons include stretch-receptor neurons that are excitatory (SRE) and inhibitory (SRI). **B.** The segmental component in the squares in Fig 23A illustrated in standard engineering form. Subscripts L and R indicate the left and right sides.

The network composed of two lateral (L) and two inhibitory (I) cells in Figs 23A and 23B is a JK flip-flop. Besides the addition of set and reset cells (E) that generate oscillations, the only difference between the lamprey's segmental component as shown in Fig 23B and the JK flip-flop of Fig 22A is an enabling signal from each input cell (E) to the corresponding output cell (I). The selective pressure that led to the lamprey's deviation from the modern JK flip-flop design is not clear, but there are at least two possibilities. With excitatory input, the output cells do not need to be spontaneously active. Spontaneously active cells may have been unavailable or uncommon in the early evolution of vertebrate CPGs. Second, electronic JK flip-flops are virtually always enabled by input from a clock. The lamprey's enabling signals from the input cells may substitute for this timing function.

### 4.8.3. Simulation of the lamprey segmental CPG

Fig 24 shows a simulation of the CPG segmental component in Fig 23B implemented with neurons.

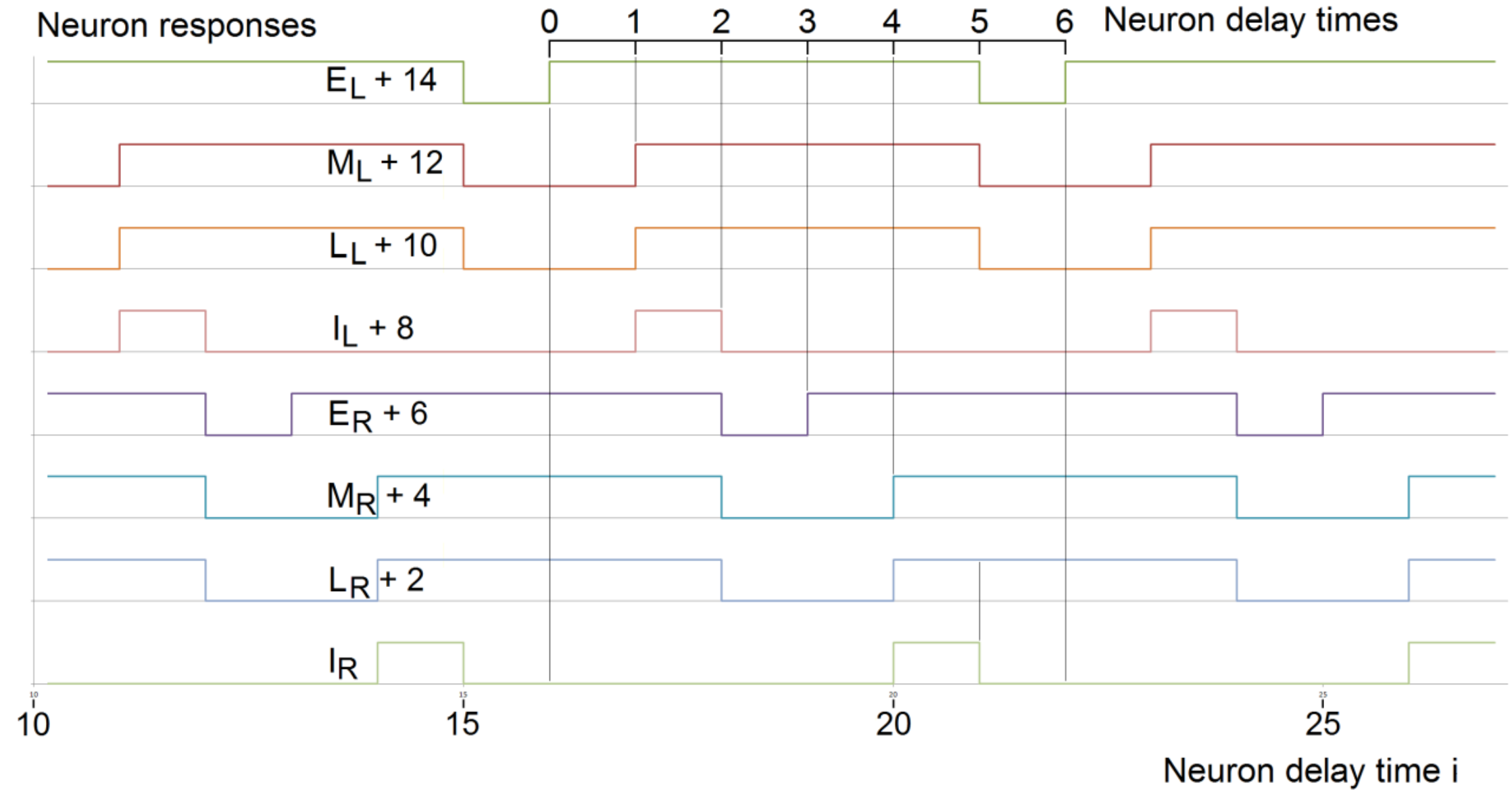

**Fig 24. Simulation of the lamprey's CPG segmental component in Fig 23B [10].** The graphs show that the common period of the neurons' simulated oscillations is six neuron delay times. The two sets of neuron responses in the left and right sides of the segment have the same oscillations, 180 degrees out of phase. This shows what produces the lamprey’s undulatory swimming motion. The motor (M) and lateral (L) neurons on each side have identical graphs because they have the same inputs, as shown in Fig 23. The graphs show the multiple state changes after each delay time, indicated by the black lines.

# 5. Conclusions

## 5.1. A framework for discovering neuron connectivity

The “form follows function” approach provides a framework for discovering neuron connectivity. Designing a simple logic circuit that can perform a single advantageous function can result in a neuronal network that generates neural correlates of multiple phenomena. The proposed networks can produce so many major, independent phenomena that actual networks in the brain are likely to be connected as proposed, at least in principle if more complex in the details. The FFF method has been successful for color vision, olfaction, short-term memory, brain waves, lobster peristaltic action, and lamprey locomotion. One of the network designs is a family of general information processors that exhibit major features of cerebral cortex physiology and anatomy. Short-term memory in NFFs and synchronization by cascaded oscillators suggest a resolution to the longstanding controversy of whether short-term memory depends on neurons firing persistently or in brief, coordinated bursts. The JK toggle’s narrow tolerances for input errors may be related to the abnormal electrical activity in neurological disorders such as epileptic seizures. Flip-flop oscillators show how two primitive ganglia that have been studied extensively can produce lobster peristaltic action and lamprey locomotion.

## 5.2. Novel logic circuits and ideas

Some of the network designs are apparently new to engineering, filling gaps and providing improvements in well-known families of logic circuits. The fuzzy decoder (Figs 6-14) that was found in [3] and refined and extended in [4] is radically different from, and an improvement on, standard Boolean decoder designs. The flip-flop ring oscillators in Fig 20 provide an even number of oscillations with uniformly distributed phases. This fills the gap in the family of inverter ring oscillators, which provide an odd number of oscillations. All four of the designs in Fig 20 are apparently new to both neuroscience and engineering. The lamprey's oscillator design in Fig 23B that controls its swimming motion is apparently new to engineering.

This is apparently the first discovery of how a known network of neurons functions as a logic circuit, and this network incorporates the most-used flip-flop design in electronic systems.

The FFF method led to several novel ideas. Brain waves resulting from synchronization of state changes to avoid timing errors was apparently first proposed in [8]. The dual transform between neuronal networks and electronic logic circuits of moving the negation symbol from one end of a connection to the other is apparently new. Any neuronal network containing AND-NOT gates that is derived from the dual transform is likely to be new to engineering as well as neuroscience because the AND-NOT gate is virtually never used as a building block in electronic logic circuit design.

The new model of color vision presented in Figs 10A and B may be a more realistic possibility for color vision than the original version in Figs 6 and 7 from [4]. The explanation of why decoders exhibit major features of the cerebral cortex (section 4.4.1.) was not included in [5].

**5.3. Predictions**

Besides producing many known phenomena, the network designs produce testable predictions of unknown phenomena, too many to include here. The best example may be the hypothesis that neural structures synchronized by cascaded oscillators can generate the exact distribution of brain wave frequencies. Although the hypothesis is consistent with available data, as illustrated in Fig 19, the data are too imprecise for a rigorous statistical test. The hypothesis can be rigorously tested by easily obtained random samples and standard tests for equal means and variances. Normal distributions are completely determined by their means and variances. The means and variances of both neuron delay times and periods of one or more EEG brain wave bands can be estimated from random samples. The EEG sample statistics can then be compared to the EEG parameters that cascaded oscillators predict from the delay time statistics. This test of the hypothesis is described in more detail in [8].

## 5.4. Observations

Some simple observations that led to the FFF method were apparently first noted in the FFF papers. The observations that excitation and inhibition can perform logic operations, that a neuron can function as an AND-NOT logic gate, that the gate is functionally complete, and that inhibition can play a role in logic and information processing were apparently new. Standard neuroscience textbooks still do not include any of these possibilities.

Other observations include: Color vision is a decoder; the unique spectral colors blue, green, and yellow occur at or near the intersections of the cone sensitivity curves; and the wavelengths of the unique spectral colors are bounds on the spectra of perceived colors. The relationship between mutually exclusive colors and highest and lowest cone type absorption follows from a comparison of the visible spectrum to the cones' sensitivity curves. The dual transform to derive Fig 15E was first observed for the standard electronic flip-flop in Fig 15D. The discovery that the extended decoders exhibit several of the major features of the cerebral cortex was simply observed.

The architectural maxim "form follows function" is itself an observation for living organisms, and it is not a new idea in biology. Except for the FFF papers, however, this simple observation apparently has not been applied in the investigation of neuron connectivity.

## 6. Acknowledgements

Simulations were done with MS Excel and CircuitLab. Network diagrams were created with CircuitLab and MS Paint. Graphs were created with Converge 10.0, MS Excel, and MS Paint. I would like to thank Arturo Tozzi, David Garmire, Robert Barfield, Ken Schoolland, Paul Higashi, Anna Yoder Higashi, Sheila Yoder, and especially Ernest Greene and David Burress for their support and many helpful comments and suggestions.